\documentclass[11pt,a4paper]{article}
\usepackage[T1]{fontenc}
\usepackage[utf8]{inputenc}
\usepackage{authblk}
\usepackage{longtable}
\usepackage{graphicx}
\usepackage{lscape}

\newcommand{\mnras}{MNRAS }
\newcommand{\pasp}{ASP}

\newcommand{\apj}{ApJ }

\newcommand{\aap}{A\&A }

\date{}

\title{Statistical Analysis of the Parameters of
Gamma-Ray Bursts with~Known~Redshifts and Peaked Optical Light
Curves}
\author[2]{Beskin G.$^{1,}$}
\author[3]{Oganesyan G.}
\author[4]{Greco G.}
\author[2]{Karpov S.$^{1,}$}
\affil[1]{Special Astrophysical Observatory of the Russian AS, Nizhnij Arkhyz, 369167 Russia; beskin@sao.ru}
\affil[2]{Kazan Federal University,Kazan, 420008 Russia }
\affil[3]{Southern Federal University, Rostov-on-Don, 344000  Russia}
\affil[4]{University of Urbino, Urbino, 61029 Italy}

\usepackage[left =2cm, right = 2cm, top = 3cm, bottom = 3cm]{geometry}
\begin{document}
  \maketitle
  
\begin{abstract}

We present the  statistical analysis of the properties of
gamma-ray bursts with measured  host galaxy  redshifts and  peaked
optical light curves in proper  frames of reference. The optical
transients are classified by comparing  the time lag of the
optical peak relative to  the GRB trigger  with the duration of
the gamma-ray emission itself. The results of the correlation
analysis of all possible pairs of energy, spectral, and temporal
characteristics of both  gamma-ray  and optical emissions are
given. We specify the pairs of the parameters with correlation
coefficients greater than 50\% at significance levels better than 1\%.
The following empirical relations, obtained for the first time,
are specifically discussed: a correlation between the peak optical
afterglow $R$~band luminosity and redshift  $L_{R} \propto
(z+1)^{5.39 \pm 0.74}$ and a correlation between the peak
luminosity of the prompt optical emissions and the time of the
peak $L_{R} \propto T_{\rm peak}^{-3.85 \pm 1.22}$. We also
analyze the similarity of the relationships between the peak
optical luminosity and the isotropic equivalent of the total
energy of gamma-ray bursts for afterglows   ($L_{R} \propto E_{\rm
iso}^{1.06 \pm 0.22}$) and for prompt optical emissions ($L_{R}
\propto E_{\rm iso}^{1.59 \pm 0.21}$).
\end{abstract}

\section{INTRODUCTION}

Gamma-ray bursts (GRB) are the brightest  transient phenomena in
the Universe (the isotropic equivalent of their total energy
reaches $10^{54}$~erg). Along with actual bursts, variable sources
related to them are also observed in different ranges: radio,
infrared, optical, and x-ray.  They are less intensive but
powerful enough, with luminosities reaching
$10^{51}$~erg\,s$^{-1}$. A division of all these manifestations
into two types of transients is generally accepted: emission
synchronous with gamma-ray (prompt), continuing at least until
$T_{90}$  (the time till which 90\% of the gamma-ray burst energy
is released), and the afterglows, usually starting after the burst
turns off. The most popular description of their population is a
model which combines the inner  and outer shock waves produced
while the fragments of the relativistic outflow decelerate due to
the collisions inside of it (prompt) and while it decelerates
in the outer interstellar medium
(afterglow)~\cite{meszaros_97:Beskin_n,rees_94:Beskin_n}.

Most of the studies are devoted to examining the properties of
individual gamma-ray bursts and to attempts of verification of
phenomenon models in general. As time has shown, in this approach
its self-consistent physical picture is far from being
constructed, especially for the optical components of the bursts.
It is still unclear whether there occur reverse shock waves, what
energy structure does a relativistic jet have, what is the density
distribution in the ambient interstellar medium surrounding the
burst, etc. In these circumstances, the search for statistical
relationships between various parameters of bursts in their
subsamples formed according to various criteria and with the
minimal use of model representations remains a topical task. In
its context, a particularly important role has the analysis of the
properties of the sample of gamma-ray bursts with measured
redshifts, numbering about 250 objects~\cite{gehrels:Beskin_n}.
This  amount of data allows us to perform a statistical analysis
of the parameters of  gamma-ray bursts in the proper frame of
reference, revealing empirical relations  between  intrinsic physical
characteristics~\cite{norris:Beskin_n,
borgonovo:Beskin_n, amati:Beskin_n, guidorzi:Beskin_n,
ghirlanda_05:Beskin_n, ghirlanda_06:Beskin_n, li_06:Beskin_n}. At
the same time, similar relations are also analyzed
  for the optical components of the bursts~\cite{zeh:Beskin_n,
nardini:Beskin_n, kann:Beskin_n, li_12:Beskin_n,
zaninoni:Beskin_n, melandri:Beskin_n}, the host
galaxies~\cite{schady:Beskin_n,zafar:Beskin_n}, and the
characteristics of the local interstellar medium  in the GRB
progenitors~\cite{schulze:Beskin_n}.

The main problem of all such studies is to find an optimum ratio
between the object selection criteria and the sample size: the
latter decreases for a more rigorous selection, and subsequently
the significance of established statistical dependences falls,
while the role of selection effects increases.
 We study the correlations
between the  parameters of GRBs with measured redshifts and peaked
optical light curves. The latter condition isolates a specific
physical condition within any given model: deceleration of either
the ejecta colliding inside the jet in the case of prompt optical
emissions or the ejected matter in the outer interstellar medium.
Notice that  it is not known in advance which kind of physical
processes would determine the occurrence of this peak. The
division of objects into subsamples was carried out by comparing
the time interval between the beginning of the optical brightness
increase and  the time
 of the trigger with the duration of the  gamma-ray emission itself (prompt
optical emissions, afterglows, and afterglows accompanied by
residual gamma-ray radiation), while the identification of the
physical type of optical emissions (prompt or afterglow) stems
from the statistical analysis of the entire subsample. This way,
some objects, initially classified as the prompt optical emissions
were subsequently found to be close to the afterglows by their
statistical properties. Notice that the use of a sample of GRBs
with known redshifts for the statistical analysis allows us to
pose the problem of the cosmological evolution of  the properties
of both the bursts themselves and the interstellar medium in the
region of their localization. Section~2 of this paper gives the
definitions and techniques of obtaining  different characteristics
of the objects, and describes the methods of statistical analysis
of the initial data and the obtained results. Sections~3,~4,~5
discuss  some of the most important correlations between the GRB
parameters we found: the relationship between the optical
luminosity at the light curve peaks
 and the redshift, between the optical luminosity and
the burst energy in the gamma-ray range, and between the
luminosity and the time of the peak. Section~6 briefly summarizes
the results obtained.

\section{THE GRB SAMPLE}

Fifty-four gamma-ray bursts  with   known redshifts and peaked
optical light curves were selected for the analysis. The time
dependences of the detected radiation flux both in the gamma-ray
and optical (in stellar magnitudes) ranges served as the data
sources for determining  GRB parameters. In the latter case, these
are, as a rule, the light curves in the $R$~band, $R(t)$,
sometimes with modified filters,  $r'(t)$, $R_{C}(t)$, and $r(t)$,
in one
  case in the $V$~band (GRB\,050730), and  in two cases in the white light
(GRB\,100906A and GRB\,080810).
With the use of   calibrations from~\cite{fukugita:Beskin_n}, spectral flux densities
 in the respective  $F_{\nu}$ bands in the observer's reference frame were found.
 The $F_{\nu}$  values  were converted into the spectral flux densities $F_{R}$
  at the effective $R$-band wavelength   in the proper  frame of reference.
   If there were no data on the  optical
spectral  index $\beta$, $F_{\nu} \propto \nu^{-\beta}$,  the mean
value \mbox{$\beta=0.75$}~\cite{panaitescu_13:Beskin_n} for GRB
optical components was used, and the following values were
determined: the flux in the observer's frame of reference
$$
F_{\rm
opt}=\int\limits_{\nu_{1}}^{\nu_{2}}F_{0}\,\nu^{-\beta}\,d\nu,
$$
where $\nu_{1}$ and $\nu_{2}$ are the frequency boundaries of a
photometric band in the observer's frame of reference; the
integral optical flux $S_{\rm opt}$  (the result of   numerical
integration of $ F_{\rm opt}(t)$ over time); the time of the
optical peak relative to the trigger time of a gamma-ray burst
$t_{\rm peak}$; the   increase and decrease indices of the optical
radiation intensity, $F_{\rm opt} \propto t^{\alpha_{r}}$,
\mbox{$F_{\rm opt} \propto t^{\alpha_{d}}$}. The parameters of the
gamma-ray emission---the peak $F_{\rm iso}$ and integral $S_{\rm
iso}$ fluxes in the range of 15--350~keV, the time of the release
of 90\% of the GRB energy $t_{90}$, and the spectral index of the
energy distribution $\alpha$---were adopted from Butler
et~al.~\cite{butler:Beskin_n}. Table~1 lists the data on the
gamma-ray bursts and their parameters in the observer's frame of
reference.
Taking into account the extinction in the
Galaxy~\cite{schlegel:Beskin_n}, the brightness of the host galaxy
(if available), and the extinction in it,  the following GRB
parameters were determined for the standard cosmological model
with $\Omega_{M}=0.3$, $\Omega_{\Lambda}=0.7$,
\mbox{$H_{0}=70$~km\,s$^{-1}$\,Mpc$^{-1}$} in the proper frame of
reference: the maximal optical $R$~band luminosity
\begin{equation}\nonumber
L_{R}=4 \pi D^2
\int\limits_{\nu_{R_{0}}}^{\nu_{R_{1}}}F_{0}\,\nu^{-\beta}\,d\nu,
\end{equation}
where $\nu_{R_{0}}$ and $\nu_{R_{1}}$ are the frequency boundaries
of the standard $R$~band, and $D$ is the photometric distance; the
total optical energy  $E_R$ for the same range (the time integral
of $L_R(t)$); the time-related parameters \mbox{$T_{\rm peak}$,
$T_{90}$} taking into account cosmological contraction. The
isotropic equivalent of the  gamma-ray burst  energy in the
extended range of \mbox{1--10\,000~keV} $E_{\rm iso}$ was adopted
from~\cite{butler:Beskin_n}, the peak luminosity $L_{\rm iso}$ was
derived from   $F_{\rm iso}$ as
\begin{equation}\nonumber
L_{\rm iso}=4 \pi D^2 F_{\rm iso}.
\end{equation}

If the extinction $A_R$ in the host galaxy  was not measured, then
based on the data on $A_V$ for 76~bursts from~\cite{kann:Beskin_n}, we determined the average of
these values at a corresponding redshift and transformed them into
$A_R$ using the model of extinction in the Magellanic
Clouds~\cite{pei:Beskin_n}. The data on the parameters of the
gamma-ray bursts in the proper frame of reference are presented in
Table~2.

All the optical transients have been classified into one of four
 following types based on the ratio between the duration
of gamma-ray emission itself and the lag of the beginning of the
optical brightness increase relative to the time of the trigger.
\begin{list}{}{
\setlength\leftmargin{2mm} \setlength\topsep{2mm}
\setlength\parsep{0mm} \setlength\itemsep{2mm} }
 \item  $\bullet$ P (prompt); thirteen  optical transients are classified
 here whose brightness begins to  increase before   the attenuation of
gamma-ray radiation. This type includes four events, referred to
as P?, whose belonging to P is ambiguous; they may belong to the
A(U) type (see below).
 \item  $\bullet$ A (afterglow); 34~transients  whose
brightness begins to increase after   the gamma-ray activity  is
ended.
 \item $\bullet$ A(U; seven transients whose
brightness is increasing in the presence of underlying residual gamma-ray
radiation after 90\%  of the total  burst energy is released.
\end{list}

For the statistical analysis, seven samples were formed
 containing the following types of objects:
\begin{list}{}{
\setlength\leftmargin{2mm} \setlength\topsep{2mm}
\setlength\parsep{0mm} \setlength\itemsep{2mm} }
 \item (1) P and P?-type transients;
 \item (2) only afterglows A;
 \item (3) only the transients with underlying continuing gamma-ray activity  A(U);
 \item (4) all 54 optical transients;
 \item (5) afterglows A and transients with underlying continuing gamma-ray activity  A(U);
 \item (6) only prompt optical emissions P;
 \item (7) the A-type transients along with A(U) and P?.
 \end{list}
Unweighted Pearson correlation coefficients   $R$ and their
significance levels  ${\rm SL}$ were
 determined between all possible pairs of  GRB  parameters
$x$ and $y$ from these samples  by the formulas
\begin{equation}
R(x,y) = \displaystyle{\frac{\sum
(x_{i}-\bar{x})(y_{i}-\bar{y})}{\sqrt{\sum (x_{i}-\bar{x})^{2}
\sum (y_{i}-\bar{y})^{2} }}} \nonumber,
\end{equation}
\begin{equation}
{\rm SL}(R)=1- \displaystyle{\frac{2 \Gamma
\left(\displaystyle{\frac{n-1}{2}}\right)}{\sqrt{\pi}\, \Gamma
\left({\displaystyle{\frac{n-2}{2}}}\right)}}\,
\int\limits_{0}^{R} \left(1-r^{2}\right)^{\frac{n-4}{2}} dr
\nonumber,
\end{equation}
where $\bar{x}$ and $\bar{y}$ are the average values of  $x$ and
$y$, $\Gamma$ is the Euler gamma function, $n$ is  the number of
points $(x_{i}, y_{i})$. Correlations between the parameters with
\mbox {$R>0.50$} and the significance  levels of \mbox{${\rm
SL}<1\%$} were assumed to be significant, and the linear regression
coefficients were then determined for these parameters  (Table~3).

\section{$L_{\rm opt}:(z+1)$ CORRELATION}

\begin{figure}[ht!]

 \includegraphics[width=1.\textwidth]{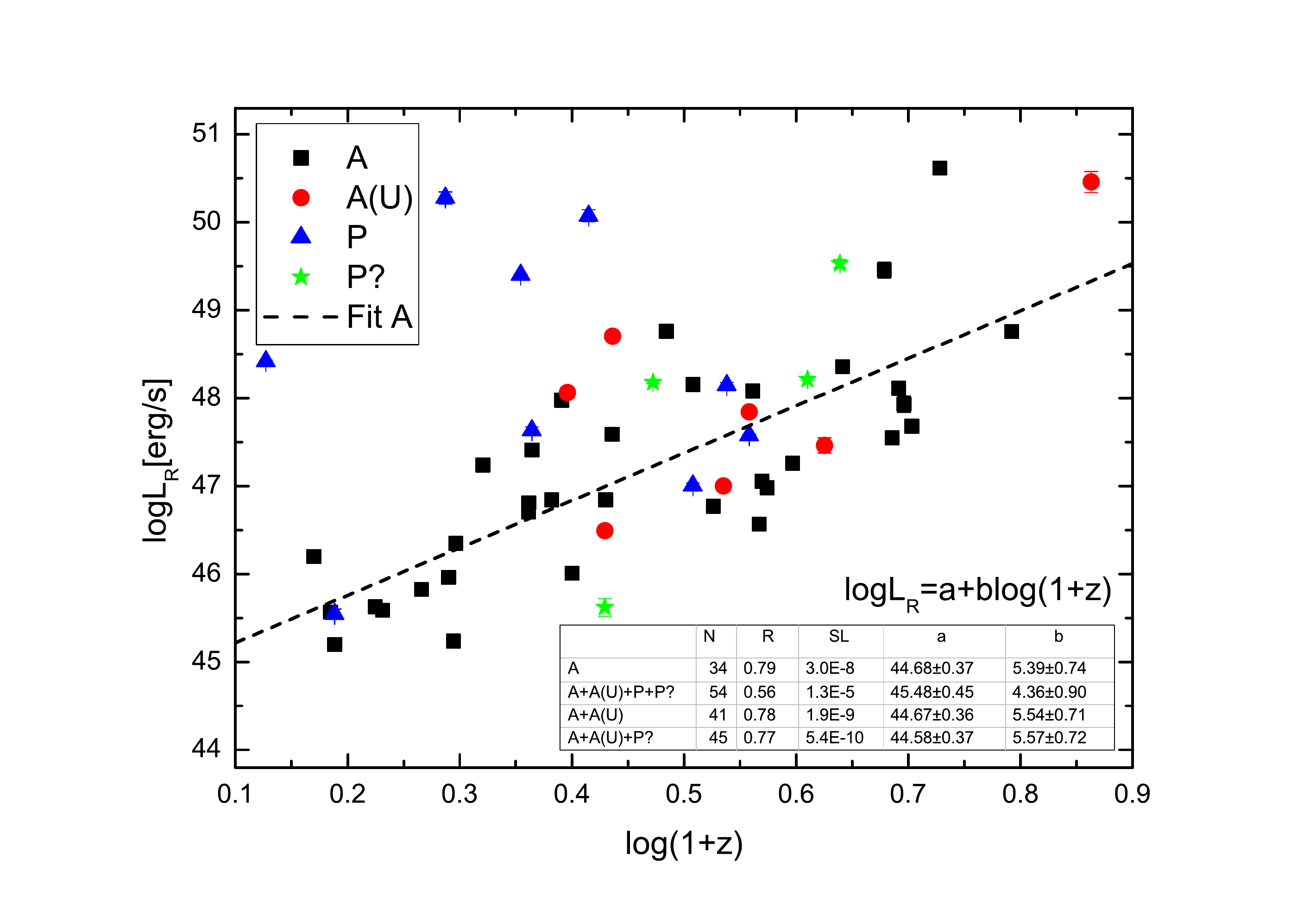}
 \caption{Dependence of the peak optical  $R$ band luminosity  $L_{R}$ on  redshift $z$.}
  \label{fig1:Beskin_n}
\end{figure}

\begin{figure}[ht!]
 \includegraphics[width=1.\textwidth]{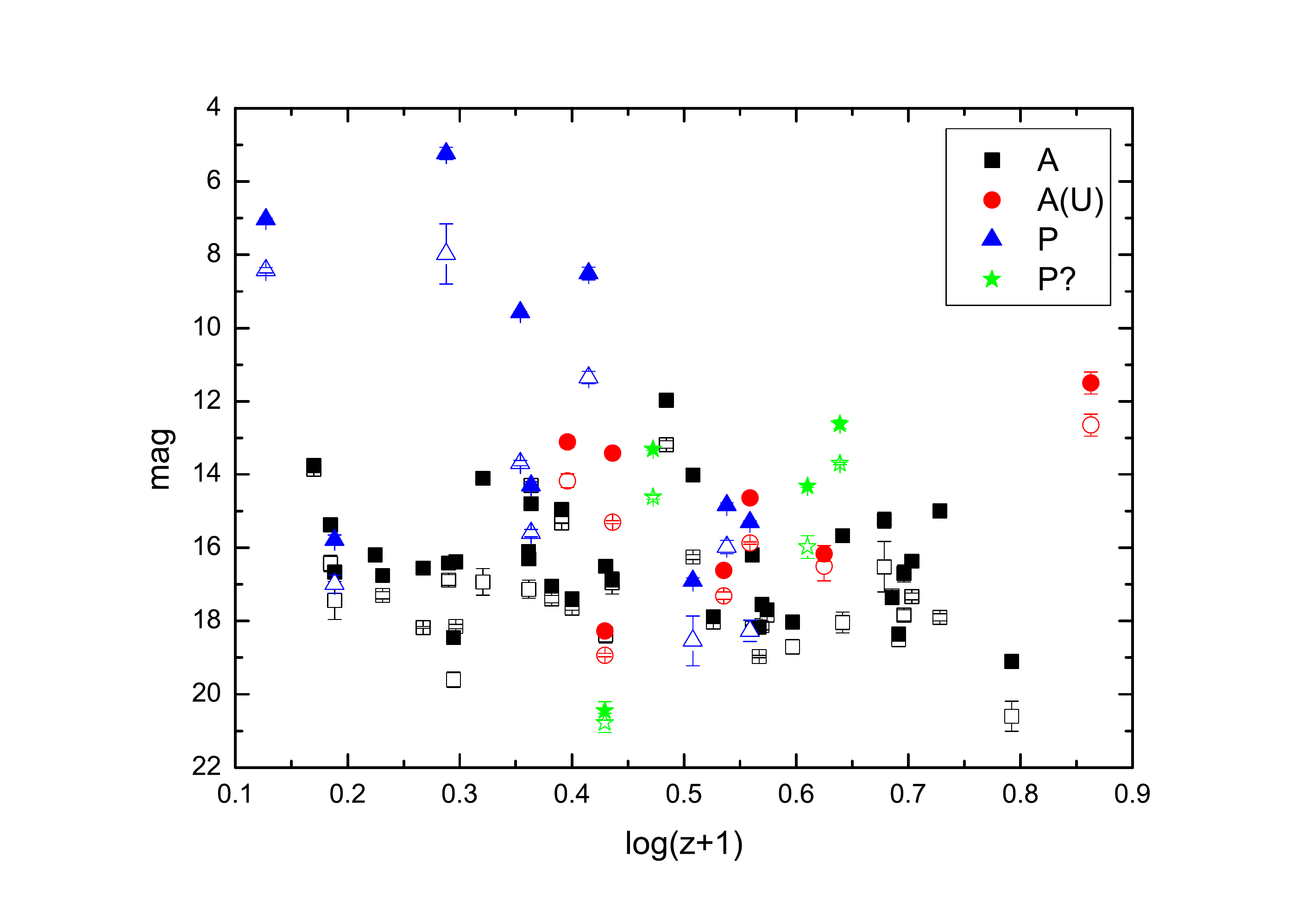}
  \caption{Observed magnitudes of the optical transients (in arbitrary filters) at the
moments of discovery (empty symbols) and at the peaks of the light
curves (filled symbols)  depending on  redshift $z$.}
    \label{fig2:Beskin_n}
\end{figure}

Among all the correlation relationships established in this study,
the $L_{R}:(z+1)$ dependence (Fig.~\ref{fig1:Beskin_n}),
discovered for the first time, in our opinion presents the
greatest interest. In particular, its presence leads to the
concept of probable cosmological evolution of the local
interstellar medium in the gamma-ray burst birthplaces. Notice
that this correlation is highly significant for all object types
except for the P-type sample objects, which is corroborating
evidence that \mbox{P?-type} transients belong to afterglows (A
and \mbox{A(U)-type} objects, whose characteristics virtually
coincide in the discussed correlation). To be completely confident
that the discovered correlation is real, the possible role of selection
effects in its occurrence has to be analyzed. Notice that to
determine the effect of selection  on the parameters of the bursts
in the samples of different types, it is first of all necessary to
examine the whole set of   brightness estimates for optical
companions of the bursts at the time of their discovery. To this
end, we built the dependences   of the observed magnitudes on
$(z+1)$ of optical  companions at the times of their discovery and
the light curve peak (Fig.~\ref{fig2:Beskin_n}). Based on the data
of Figs.~1 and~2, the following conclusions can be formulated.
Given a relatively equal luminosity at different redshifts in the
proper reference frames, the observed flux should drop
quadratically with  $(z+1)$,  and therewith at maximum redshifts
reach the typical detection limit at around $23^{\rm m}$,
estimated in our sample from the total light curves. This effect
would lead to a lack of low-luminosity objects in the proper frame
of reference, whose observed brightness is fainter than $23^{\rm
m}$, and as a consequence, to a luminosity increase with
 growing redshift. Nevertheless, no regular radiation flux
decrease is observed either in the whole set of brightness
estimates for optical transients at the moment of their detection
or in the peak brightness. On the other hand, both at $z>4$ and
$z<2$ the brightness of objects varies within close limits from
$19^{\rm m}$--$20^{\rm m}$ to $13^{\rm m}$--$14^{\rm m}$, by far
exceeding the detection limit of $23^{\rm m}$.
 In other words, the  selection by brightness is absent in our data.
 
We compared the correlation coefficients and linear regression
parameters for the group of transients from the A + A(U) + P?
sample,   localized in different  redshift ranges (Table~4). It is
easy to see that even if we exclude the  transients with  $z>4.5$
and $z<1$, the luminosity increase with increasing  redshift
remains detectable and significant. Clearly, if we  narrow the
redshift range,  the  correlation coefficients decrease, but
 nevertheless the linear regression parameters for different
groups of transients  remain the same within their
errors; in particular,  the exponents are close to    4--5.

\begin{table}
\centering
 \caption{Characteristics of the redshift dependence of the A\,+\,A(U)\,+\,P?  sample peak optical luminosity  in different $z$
ranges: the number of transients in the sample $N$, the Pearson
correlation coefficient $R$, the \mbox{$\log L_R = a+b \log
(z+1)$} linear regression parameters~$a, b$}
\medskip
\label{table:2:Beskin_n}
\begin{tabular}{c|c|c|c|c|c}
 \hline
  $z$     & $N$ & $R$ & ${\rm SL}$ & $a$ & $b$ \\
\hline
all               & 45 & 0.77 & $0.54\!\times\!10^{-9}$ & $44.58\!\pm\!0.37$ & $5.57\!\pm\!0.72$ \\
$z\!<\!4.5$       & 43 & 0.72 & $0.65\!\times\!10^{-7}$ & $44.67\!\pm\!0.40$ & $5.55\!\pm\!0.81$ \\
$z\!>\!1$         & 36 & 0.61 & $0.84\!\times\!10^{-4}$ & $45.08\!\pm\!0.62$ & $4.96\!\pm\!1.11$ \\
$1\!<\!z\!<\!4.5$ & 34 & 0.50 & $0.23\!\times\!10^{-2}$ & $45.39\!\pm\!0.70$ & $4.31\!\pm\!1.30$ \\
\hline
\end{tabular}
\end{table}

Thus, this brief analysis shows that the observed
  correlation may not be due to the effects of observational
selection of optical transients by their brightness.

\begin{figure}[ht!]
 \includegraphics[width=1.\textwidth]{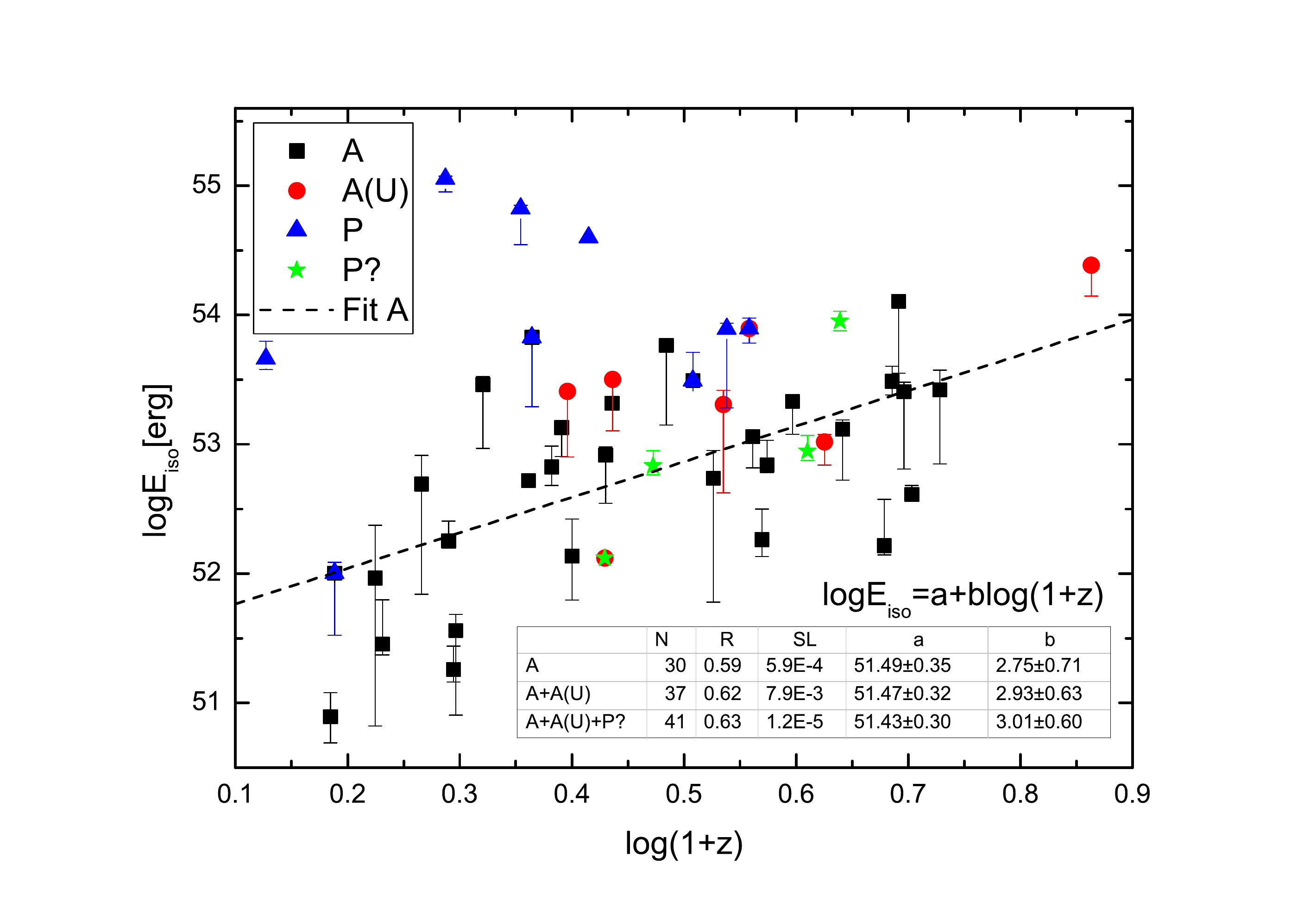}
\caption{Dependence of $E_{\rm iso}$ on redshift $z$.}
  \label{fig3:Beskin_n}
\end{figure}

\begin{figure}[ht!]
 \includegraphics[width=1. \textwidth]{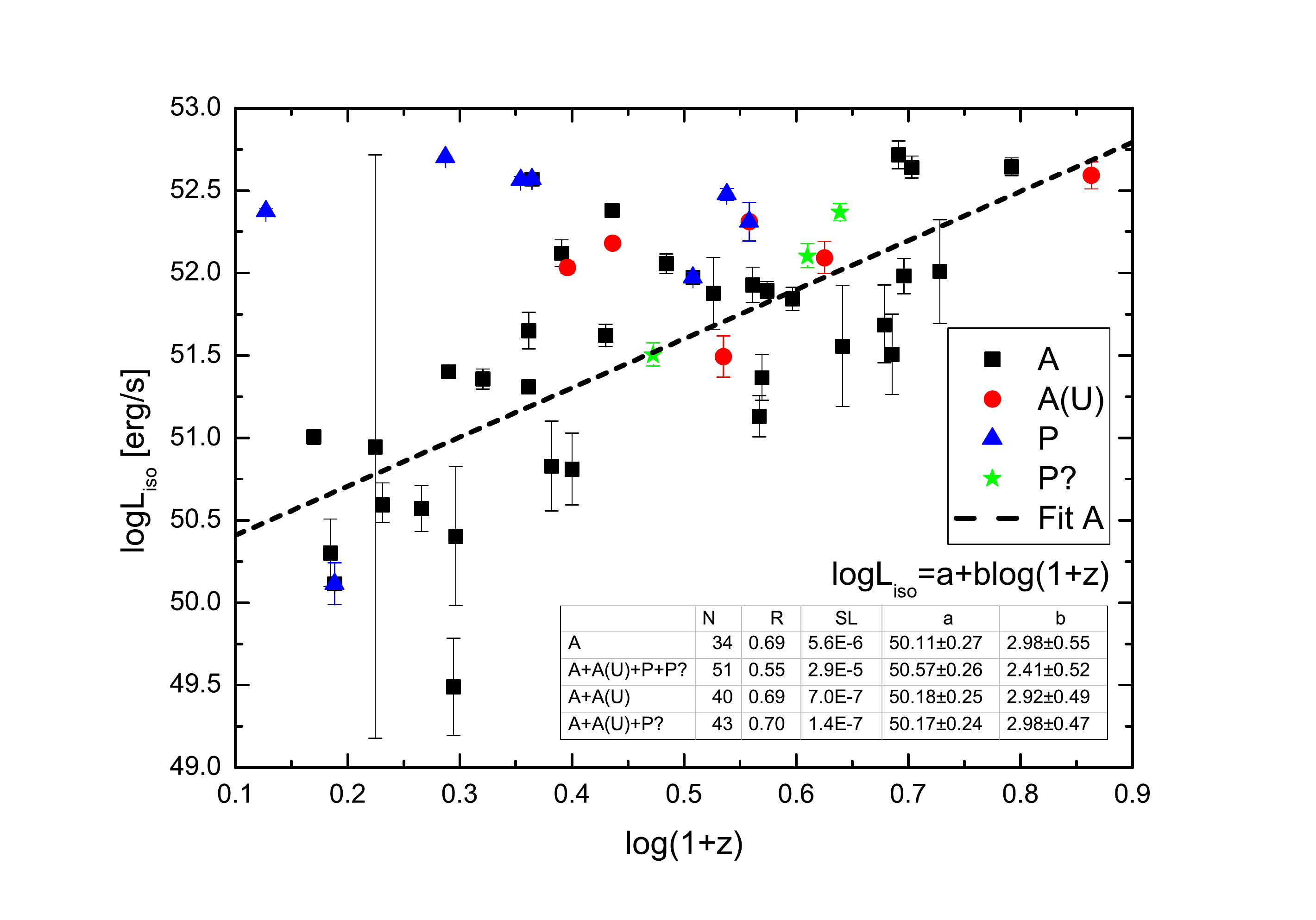}
 \vspace{1mm}
  \caption{Dependence of $L_{\rm iso}$ on redshift $z$.}
  \label{fig4:Beskin_n}
\end{figure}

To determine the statistical significance of the discovered
correlation, we generated a sample of 100\,000~optical
transients normally distributed by  $L_{R}$  with the average
\mbox {$\log L_{R}=46$}, $\sigma=2$
 in the logarithmic
scale~\cite{coward:Beskin_n} (i.e., with luminosities independent
of redshift). The extinctions (Galactic and from the host galaxy)
were taken to be \mbox{$0<A_{R}<3$},   and the optical spectrum
indices   $0.2<\beta<1.2$. The $A_{R}$ and $\beta$ limits were
adopted  from the observations~\cite{kann:Beskin_n}, the
distribution was taken to be
  uniform --- the changes of distribution functions had little effect on
the result.
 Redshifts   $z$  in the range of $0<z<6$ were randomly assigned   to the transients.
  From this general population we randomly
  selected subsamples of 50~transients each (2000~subsamples) with the
observed brightness exceeding $23^{\rm m}$, determined using the
luminosity and other above-mentioned parameters.
 The coefficient of correlation between luminosity and redshift was calculated for each sample.
 Its value exceeded the 0.7  level only in two cases; in other words, the probability of type~I error (of finding a correlation in a random sample
of transients when it doesn't exist  in the general population)
is $10^{-3}$.  To estimate the probability of error of the second kind,
a general population was generated which differed from the
previous one by the set law    $L_{R} \propto (1+z)^{4}$. The
correlation coefficient proved to be less than 0.8 in only one
sub-sample, which gives the  $5 \times 10^{-4}$ for the probability of type~II error
(to miss the correlation when it exists in the
general population). Ultimately,
 the luminosity increase in  the afterglow light curve peaks with
growing GRB redshift  appears to be a real effect. It apparently
reflects the fact of cosmological evolution
 of  the interstellar medium density in the regions of GRB birth~\cite{beskin:Beskin_n}.
 Indeed, the results of the correlation and regression analysis
 (see  Fig.~1) show that the relation \mbox{$L_{R}:(z+1)$}   holds for
 all types of optical transients (samples   A, \mbox{A\,+\,A(U),}
\mbox{A\,+\,A(U)\,+\,P?})  except for the prompt ones (P) (the
correlation substantially decreases  when  they are added to the
afterglows). On the other hand,  the \mbox{P?-type}  objects can
be, with a greater confidence, classified as afterglows, and not as
prompt emissions, since the characteristics of the
\mbox{A\,+\,A(U)\,+\,P?} object sample, to which they belong, are
virtually identical to those for the afterglow samples
 A and \mbox{A\,+\,A(U),} where they are missing.
 Moreover, there isAcknowledgements
 reason to believe that only four (out of nine) brightest cases of prompt
 optical emissions, which have no relation to the correlation $L_{R}:(z+1)$,
 are true prompt emissions whose optical radiation is generated
 at the stage of internal  shock wave collisions near the central engine of the
 gamma-ray burst, and is not susceptible to  the interstellar medium.
This is their difference  from the remaining five transients of
the P sample, whose \mbox{luminosity--redshift} dependence   is
almost identical to that of the afterglows. Thus, we can assume
that there is a group of gamma-ray bursts, whose prompt optical
emissions, synchronous with gamma-ray emission, are in fact
afterglows by their  nature, i.e., the result of the  interaction
of the jet with the local interstellar medium. In other words,
this synchronicity by itself is  not a sufficient condition to
interpret the prompt optical emission within the model of internal
shock waves generated in the  collisions of individual ejecta in
the jet itself. Quite strong   correlation of the energy and peak
luminosity of GRBs with redshift
 $E_{\rm iso}:(z+1)$ and $L_{\rm iso}:(z+1)$ (Figs.~\ref{fig3:Beskin_n}~and~\ref{fig4:Beskin_n})
in the objects of the same samples gives evidence for this
assumption.
 Naturally, these relationships get stronger
 if we include in the sample   five transients of prompt optical emission, whose
 correlation  and  regression properties are virtually identical to those for the
 afterglows A\,+\,A(U)\,+\,P?. Notice that this correlation is remarkably
 more significant than the weak relationship found   for the full sample of GRBs with
 measured redshifts   ($R=0.44$)~\cite{li_07:Beskin_n}, usually explained by the
 selective properties of Swift's main detector  (reduction by the observed flux).
Finally, let us point out   yet another feature of   ``true''
bursts, the brightest events whose characteristics do not depend
on redshift: these objects have no peaks in optical afterglows,
i.e., their optical brightness decreases quite monotonically after
the peak, coinciding with the gamma-ray burst.
 This behavior can be interpreted in the context of the above
reasons as the attenuation of the radiation emitted by the
expelled jet material spreading in the space where a powerful
explosion of the central engine has swept out the interstellar
gas.

With great probability the   \mbox{$L_{R}\!:\!(z\!+\!1)$}
correlation  for afterglows is related  to the evolution of the
density of the local interstellar medium, the regions of active
star formation in the host galaxies of gamma-ray bursts, on scales
of $0<z<6$. The density of   interstellar medium determines the
character of deceleration of the front shock wave responsible for
the late optical emission, namely,  the luminosity in the denser
media is higher~\cite{gou:Beskin_n}.

The absence of peaked afterglows in some of powerful GRBs
accompanied by the prompt optical emission  is, apparently,
evidence of certain differences in their physical natures compared
to the other events, especially as, their luminosities do not
correlate with redshifts (they drop out of the discovered
dependencies). In combination with the fact that the other prompt
optical emissions (two orders of magnitude lower by the isotropic
equivalent of optical energy $E_{R}$) do not alter the
correlations obtained, this leads to the idea of specific
high-energy sources of gamma-ray bursts, not ``feeling'' the
interstellar medium. On the other hand, it can be
  assumed that there are sometimes prompt optical emissions and even gamma-ray emissions\,(!), susceptible to
the interstellar medium during and shortly after the
explosion of the central engine.

\section{ $L_{R}:E_{\rm iso}$ CORRELATION}

\begin{figure}[ht!]

 \includegraphics[width=1.\textwidth]{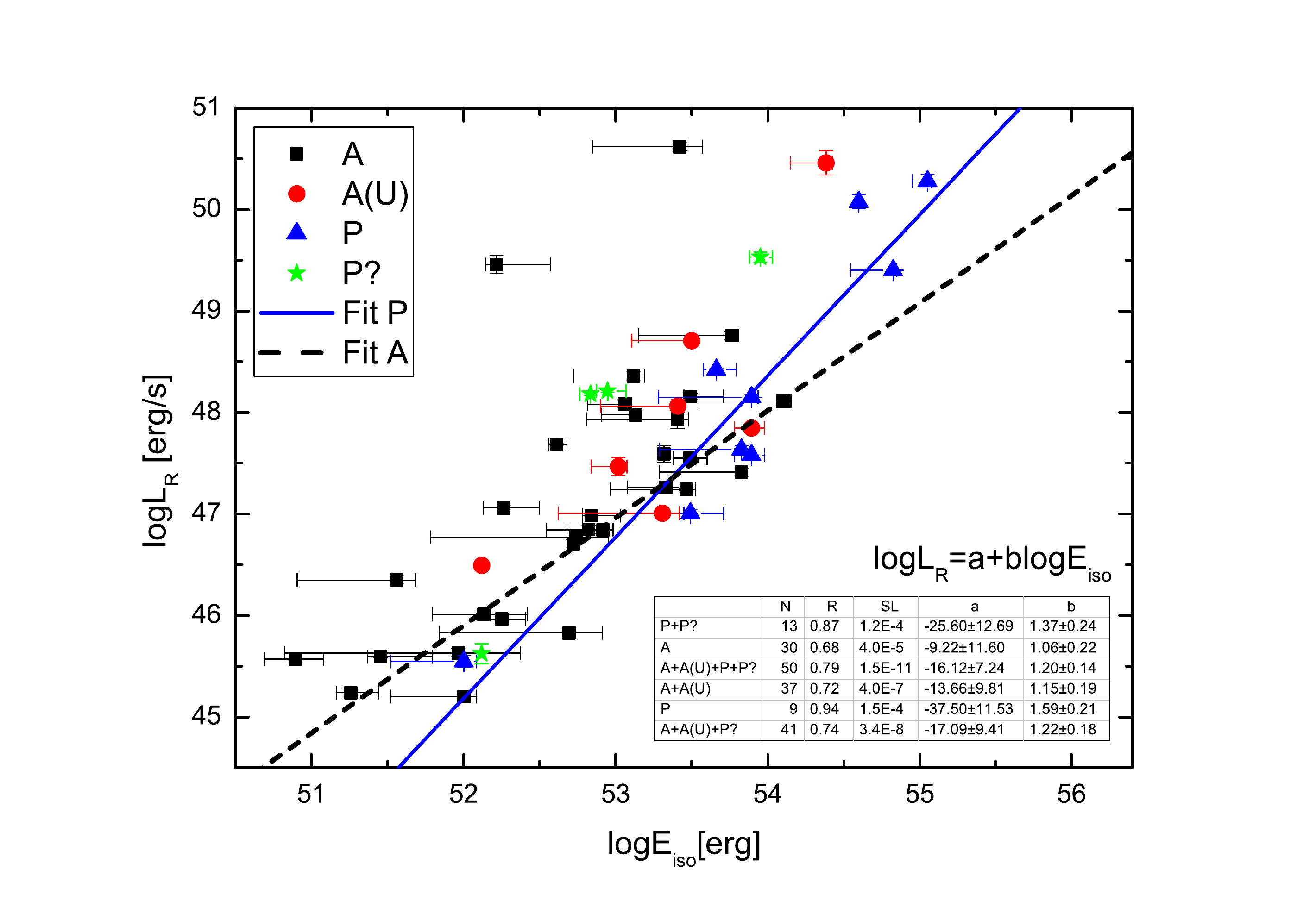}
  \caption{Dependence of the $R$~band peak optical luminosity $L_{R}$ on  $E_{\rm iso}$.}
  \label{fig5:Beskin_n}
\end{figure}

We have found the correlation between the luminosity at the
optical brightness peak and the energy of gamma-ray radiation
almost for all the transient samples (Fig.~\ref{fig5:Beskin_n}).
In particular, for the afterglows   (A)  it has the form of
\mbox{$L_{R} \propto E_{\rm iso}^{1.06 \pm 0.22}$},
 almost coinciding with the dependence    \mbox
{$L_{R}^{\rm peak} \propto E_{\rm iso}^{1.00 \pm 0.14}$}, earlier
obtained in~\cite{liang:Beskin_n}. Such a ratio is typical in the
models of deceleration of the front   shock wave in the
interstellar medium. Let us analyze to what   variant of this
process  does  our result correspond.

Consider two regions of the synchrotron spectrum in which the peak
of the optical brightness curve can be
located~\cite{panaitescu_09:Beskin_n}. If the characteristic
radiation frequency  is lower than the characteristic electron
cooling frequency $\nu_{i}<\nu_{c}$,  then the optical spectral
index $\beta=\displaystyle \frac{p-1}{2}$, where $p$ is the
spectral index of emitting electrons   $N \propto \epsilon^{-p}$,
$2<p<3$~\cite{piran_04:Beskin_n}, and $0.5<\beta<1$; yet if
$\nu_{i}>\nu_{c}$, then $\beta=\displaystyle \frac{p}{2}$ and
$1<\beta<1.5$.  At the same time, the observations give
$\beta<1$~\cite{zafar:Beskin_n}  with the mean $\beta=0.75$,
whence it follows that the radiation frequency is lower than the
frequency of   cooling. Let us now analyze the  $L_{\rm peak}
\propto E_{\rm  iso}^{\delta}$ correlation for this case, i.e., at
$\nu_{i}<\nu_{c}$~\cite{eerten:Beskin_n}.
 \begin{list}{}{
\setlength\leftmargin{2mm} \setlength\topsep{2mm}
\setlength\parsep{0mm} \setlength\itemsep{2mm} }
 \item (1) When the   energy of both the forward and reverse shock waves is emitted in the
     optically thin plasma, $L_{\rm peak} \propto E_{\rm iso}^{ \frac{12-7k-pk}{4(3-k)} }$
         and the density profile of the interstellar medium has the form of $n \propto r^{-k}$.
\item (2) For the optically thick plasma we have:
\begin{list}{}{
\setlength\leftmargin{2mm} \setlength\topsep{2mm}
\setlength\parsep{0mm} \setlength\itemsep{2mm} }
 \item (a) in the dominance of the forward  shock wave,  $L_{\rm peak} \propto E_{\rm iso}^{
\frac{12-5k+4p-pk}{4(4-k)} }$;
 \item (b) in the dominance of the reverse shock wave, $L_{\rm peak} \propto E_{\rm iso}^{\frac{20-5k-pk}{4(4-k)}}$.
 \end{list}
 \end{list}
 To determine $\delta$,  let us use the $0<k<1.5$  and  \mbox{$2<p<3.5$} estimates obtained
in~\cite{liang:Beskin_n}

\begin{figure}[ht!]
 \includegraphics[width=1.\textwidth]{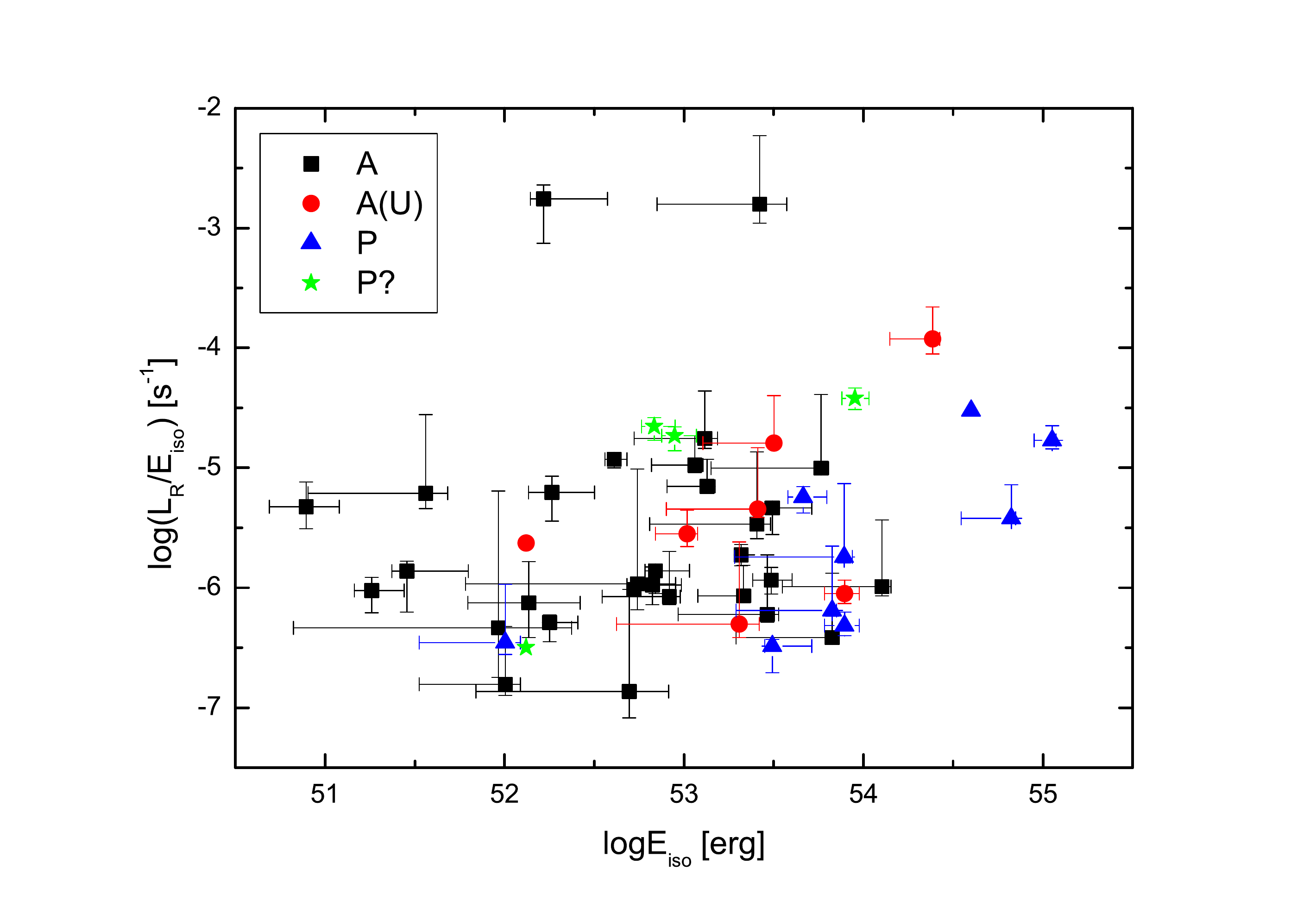}
 \caption{Dependence of the   $L_{R}/E_{\rm iso}$ ratio on  $E_{\rm iso}$.}
  \label{fig6:Beskin_n}
\end{figure}

Given the optically thin plasma, the linear dependence of
luminosity on energy  ($\delta = 1$), which is close to that we
have obtained, can be realized only by the propagation of shock
waves in a homogeneous environment, whereas the possibility of its
existence around a gamma-ray burst seems improbable. Given the
optically thick plasma,  the limits on  $\delta$ both for
the forward $0.95<\delta<1.625$ and the reverse $0.725<\delta<2.5$
shocks  are consistent with our estimate. Thus, this analysis
shows that the afterglow radiation   in the  GRB sample with
peaked optical light curves can be related both to the forward and
reverse shock waves.  Notice that the parameters of the linear
regression $L_{R}:E_{\rm iso}$  for the transients from all the
samples are very similar  within the errors (see Fig.~4). To
refine this result,  we present in Fig.~\ref{fig6:Beskin_n} the
dependence of the coefficient of transformation of
mechanical energy into optical $\kappa=L_{R}/E_{\rm iso}$ on the
gamma-ray energy. It is clear that this ratio depends both on the
actual efficiency coefficient of  the conversion of the mechanical
energy of the burst into radiation and on the opening angle of the jet.
It is easy to see that
  the behavior of this characteristic  is very similar for different
transients. It may give evidence on the existence   of a relation
between the real conversion coefficient and the opening angle of
the jet, which permits  equal $\kappa$ in the regions where the
radiation  of prompt  emissions  and
  afterglows is generated, i.e., at different distances from the
central engine.

\section{ $L_{R}:T_{\rm peak}$ CORRELATION}

The study of the character  of the correlation between the
normalized optical flux and the time of the peak relative to the
time of GRB detection  \mbox{$F_{z=z_{0}} \propto t_{\rm
peak}^{-\gamma}$} may clarify various aspects of the physical
nature of these objects. Such studies
as~\cite{panaitescu_11:Beskin_n,panaitescu_13:Beskin_n} are
devoted to the analysis of this correlation. We believe  that a
significant disadvantage of these studies is that the examined
transients were not divided by their
  possible origin: prompt  emissions, afterglows,
 and very late emissions were not identified.

\begin{figure}[ht!]
 \includegraphics[width=1.\textwidth]{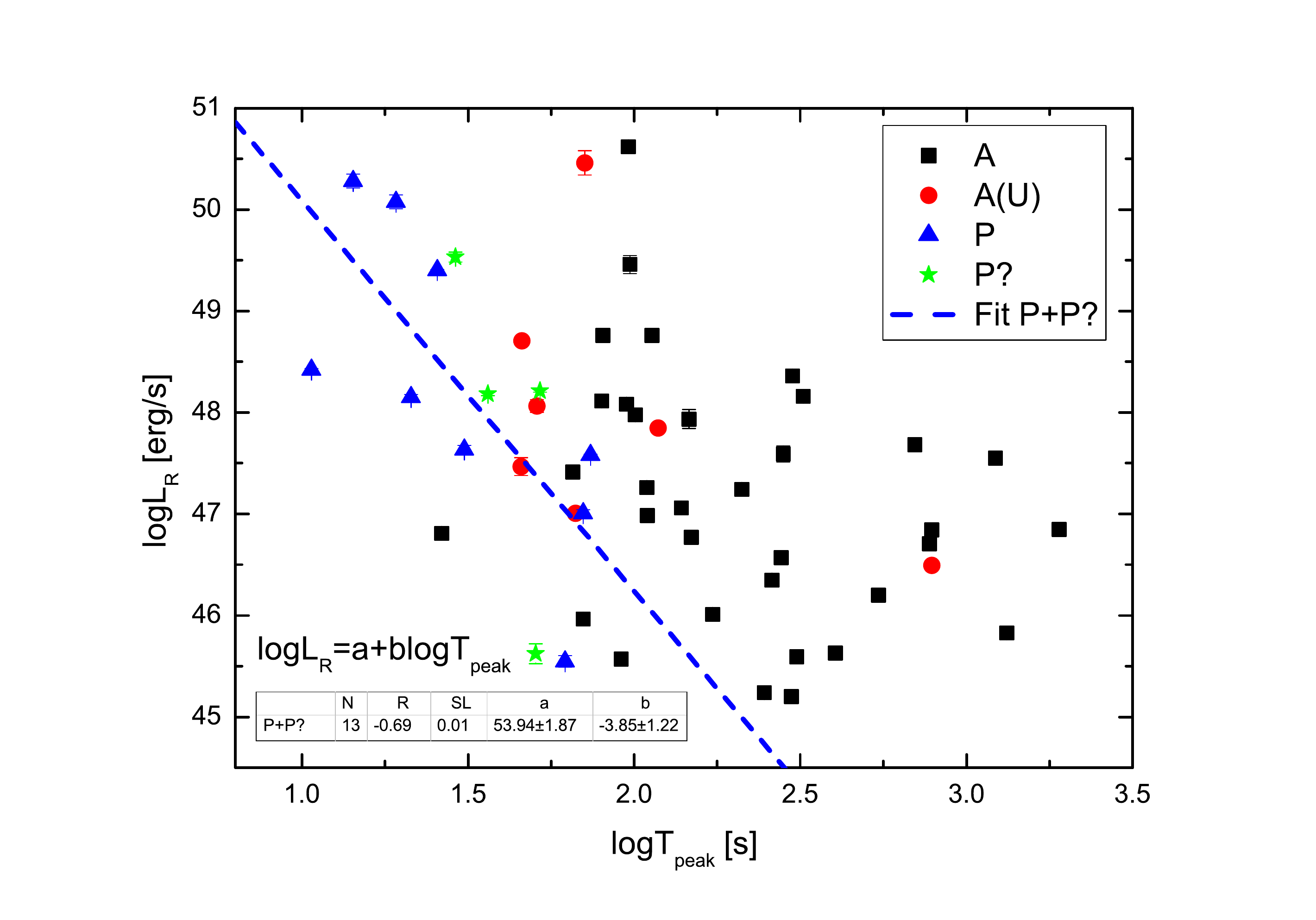}
 \caption{Dependence of the $R$~band peak optical luminosity $L_{R}$ on  the time of  the peak $T_{\rm peak}$.}
  \label{fig7:Beskin_n}
\end{figure}

\begin{figure}[ht!]
 \includegraphics[width=1. \textwidth]{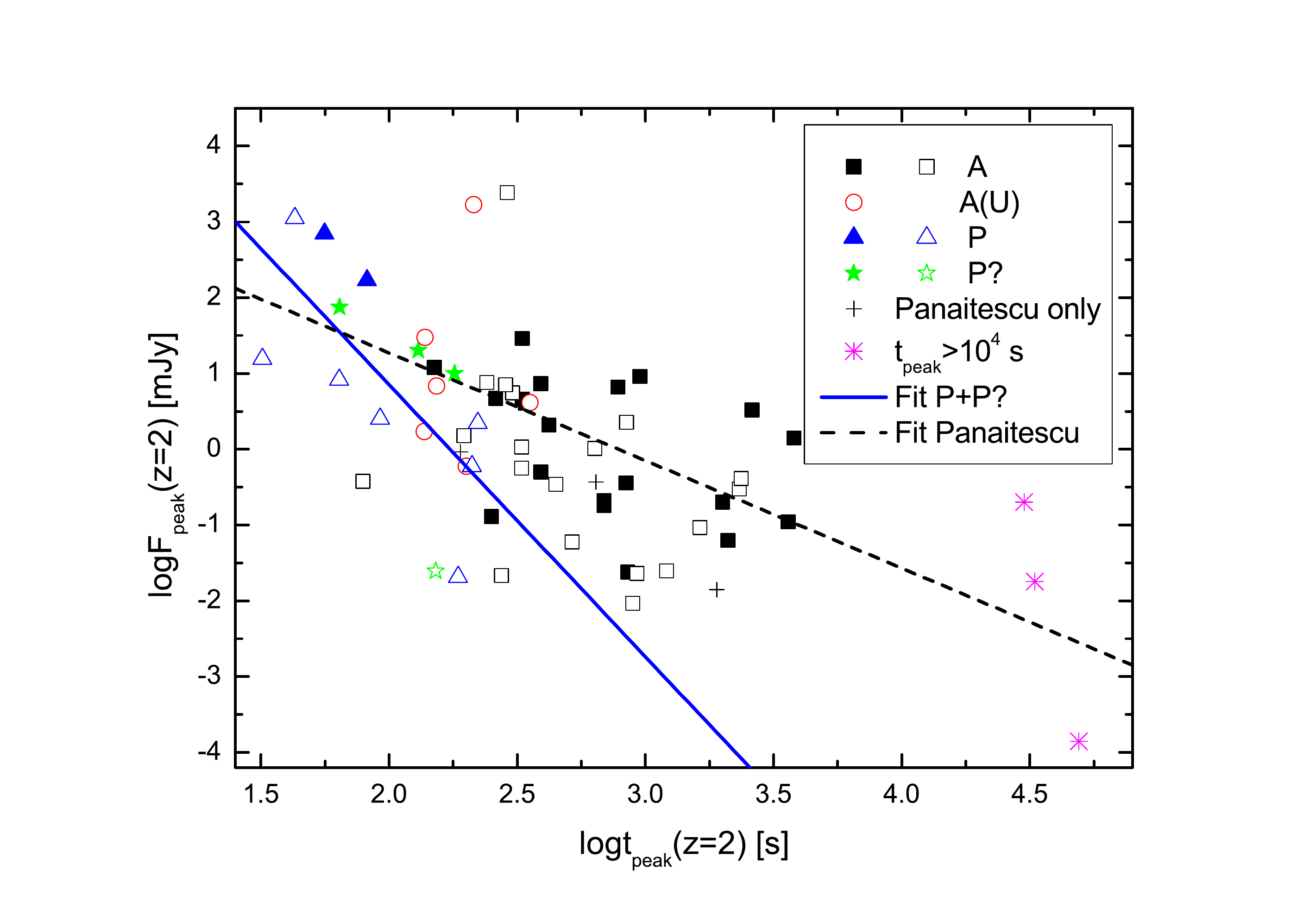}
  \caption{Dependence of the peak flux at  \mbox{$z=2$} on the time of the peak.
   The filled symbols mark the data common between our work and Panaitescu et~al.~\cite{panaitescu_13:Beskin_n}; the
  empty symbols represent the data included in our work and   absent in~\cite{panaitescu_13:Beskin_n};
 the crosses denote the data missing in our paper  and included in~\cite{panaitescu_13:Beskin_n}
  (the asterisks are the objects with \mbox{$t_{\rm peak}>10^4$~s}).}
  \label{fig8:Beskin_n}
\end{figure}
 
 In particular, in the latter case,
when the peak is reached at  \mbox{$T_{\rm peak}>10^{4}$~s},
increasing brightness on such time scales is due to the
inhomogeneities of the interstellar medium in the propagation of
the front shock wave~\cite{lazzati:Beskin_n}, in contrast to the
nature of the peaks of prompt emissions and afterglows. In our
sample, we have observed an anticorrelation between the optical
luminosity at the peak and the time of the peak; however, only for
the prompt emissions (P\,+\,P?) this correlation is characterized
by \mbox{$\gamma=3.85 \pm 1.22$} (Fig.~\ref{fig7:Beskin_n}). The
studies~\cite{panaitescu_11:Beskin_n}
and~\cite{panaitescu_13:Beskin_n} found that $\gamma = 2$   for
all the peaked light curves and $\gamma=1$  for the light curves
with a flat region. It appears that this result is of random
nature, caused by the presence of prompt emissions in the sample,
for which the ``luminosity--time of the peak'' relationship is
indeed valid, as well as several weak late emissions, creating the
long-base effect. Let us prove this. Figure~\ref{fig8:Beskin_n}
demonstrates the points of the \mbox{$F_{z=2}:t_{\rm peak}$}
diagram from~\cite{panaitescu_13:Beskin_n}, divided according to
our classification . The Pearson correlation coefficient for all
transients from~\cite{panaitescu_13:Beskin_n} \mbox{$R=-0.77$},
but if we exclude the P and P?-type emissions, as well as
emissions with \mbox{$t_{\rm peak}>10^4$~s} from this sample, the
correlation disappears \mbox{($R=-0.44$).} To compare our sample
with the sample of Panaitescu
et~al.~\cite{panaitescu_13:Beskin_n}, we have also included the
data that the latter work lacked,   which further strengthen  the
assertion about the absence of this correlation for afterglows.

The empirical correlation that we found between the peak
luminosity and the time  of the peak  for the prompt emissions can
be interpreted within the model of multiple collisions of plasma
ejecta~\cite{rees_94:Beskin_n}: every event of such a collision
generates a single peak, one or a few of which we observe. This
process may have a relaxing character: the subsequent  warming up
events in the collisions of the emitting plasma are less
energetic, which qualitatively explains the $L_{R} \propto T_{\rm
peak}^{-3.85 \pm 1.22}$ correlation that we have found.

\section{CONCLUSION}

The number of   gamma-ray bursts observed to date is great, but we
are still very far from getting a  thorough understanding   of the
nature of this phenomenon. In our work we attempted  to reveal new
properties of gamma-ray bursts with optical components. The
results of the study can be briefly summarized as follows.
\begin{itemize}

\item  Pair  correlations between the parameters of
gamma-ray bursts in the observed and proper reference
frames have been studied, and the statistically significant cases have
been identified.
 \item  A correlation has been found, free from
selection effects, between the peak luminosity in the standard
$R$~band and the redshift.
 \item   A step was made to rethink the previously found
 correlation between the peak luminosity and the time of the
 peak in the optical light curve: we demonstrate that
 such a correlation  is typical only for prompt
optical emissions.
 \item A hypothesis has been  put forward on the
 presence of a correlation
 between the  efficiency  of conversion of
the mechanical energy  of the relativistic jet  into optical
radiation and the jet opening angle, providing the similarity of
the correlations of optical luminosity and gamma-ray  energy for
sources differently localized in the jet.
\end{itemize}

\section*{Acknowledgments}
The work was   partially supported by the grant  allocated under
the Russian Government Program of Competitive Growth of Kazan
Federal University. The study of the statistical properties of
gamma-ray bursts as the events contributing to the formation of
stellar-planetary associations was supported by the Russian
Science Foundation  (project 14-50-00043, the Exoplanets
program).

\begin{landscape}
\begin{longtable}{ccccccccccc}

\caption{Parameters of gamma-ray bursts in the observer's frame of reference} \label{tab:long:Beskin_n} \\

\hline
&&&&&&&&& \\
   $GRB$ & $z$ & $Sample$ & $t_{peak}$  & $logF_{opt}$ & $logS_{opt}$  & $\alpha_{r}$ & $\alpha_{d}$ & $logF_{iso}$ & $logS_{iso}$ \\
&&&&&&&&& \\
                      &        &              &    $s$                    &       $erg/s~cm^{2}$                 &   $erg/cm^2$                     &                       &                        &        $erg/s~ cm^2$               &       $erg/cm^2$  \\
&&&&&&&&&\\
\hline
&&&&&&&&&\\

  $130427A$	&	$0.340$	&	$P$	&	$14.31$		&$-8.21$	$\pm$	$0.01$	&	$-6.80$	$\pm$	$0.04$	&	$1.43$	&	$-1.69$	$\pm$	$0.27$	&	$-4.20$	$\pm$	$0.01$	&	$-3.14$	$\pm$	$0.01$	\\	

 $130831A$	&	$0.479$	&	$A$	&	$804.04$	&$-10.83$	$\pm$	$0.01$	&	$-7.96$	$\pm$	$0.01$	&	$0.58$	&	$-0.19	$	&	$-5.93^{+0.04}_{-0.04}$	&	$-5.05$	$\pm$	$0.01$	\\					
$081007$	&	$0.530$	&	$A$	&	$139.80$		&$-11.59$	$\pm$	$0.01$	&	$-8.68$	$\pm$	$0.03$	&	$1.04$	$\pm$	$0.11$	&	$-1.14$	$\pm$	$0.05$	&	$-6.74^{+0.21}_{-0.20}$	&	$-6.23$	$\pm$	$0.04$	\\

 $060729$	&	$0.543$	&	$P$	&	$95.60$		&	$-11.63$	$\pm$	$0.06$	&	$-10.17$	$\pm$	$0.08$	&	$6.72$	&	$-3.51$	$\pm$	$0.80$	&	$-6.95$	$\pm$	$0.13$	&	$-5.39^{+0.04}_{-0.03}$	\\	

 $060729$	&	$0.543$	&	$A$	&	$458.80$	&$-11.97$	$\pm$	$0.04$	&	$-8.64$	$\pm$	$0.05$	&	$0.80$	$\pm$	$0.12$	&	$-0.21$	$\pm$	$0.04$	&	$-6.95^{+0.13}_{-0.13}$	&	$-5.39^{+0.04}_{-0.03}$	\\
	
  $111209A$	&	$0.677$	&	$A$	&	$676.08$	&	$-11.82$	$\pm$	$0.04$	&	&	&	&	$-6.35$	$\pm$	$1.77$	&	$-5.88^{+0.12}_{-0.13}$	\\
										
 $060904B$	&	$0.703$	&	$A$	&	$526.00$		&	$-11.97$	$\pm$	$0.01$	&	$-7.93$	$\pm$	$0.02$	&	$0.79$	$\pm$	$0.5$	&	$-1.07$	$\pm$	$0.06$	&	$-6.74^{+0.13}_{-0.11}$	&	$-5.73^{+0.09}_{-0.07}$	\\	
	
 $080710$	&	$0.845$	&	$A$	&	$2441.00$		&	$-11.91$	&	$-8.12$	&	$1.08$	$\pm$	$0.5$	&	$-0.54$	$\pm$	$0.01$	&	$-6.97$	$\pm$	$0.14$	&	$-5.49$	$\pm$	$0.07$	\\		
 $080319B$	&	$0.937$	&	$P$	&	$27.65$		&	$-7.55$	$\pm$	$0.07$	&	$-6.08$	$\pm$	$0.04$	&	$2.85$	$\pm$	$0.54$	&	$-4.28$	$\pm$	$0.34$	&	$-4.95$	$\pm$	$0.01$	&	$-3.39$ $\pm$	 $0.01$\\

 $071010B$	&	$0.950$	&	$A$	&	$136.99$	&	$-11.88$	$\pm$	$0.04$	&	$-9.76$	$\pm$	$0.05$	&	$0.62$	$\pm$	$0.11$	&	$-0.38$	$\pm$	$0.18$	&	$-6.26^{+0.04}_{-0.03}$	&	$-5.29$	$\pm$	$0.02$	\\	

 $070419A$	&	$0.971$	&	$A$	&	$485.82$		&	$-12.56$	$\pm$	$0.05$	&	$-9.10$	$\pm$	$0.07$	&	$1.93$	$\pm$	$0.28$	&	$-0.53$	$\pm$	$0.06$	&	$-8.20$	$\pm$	$0.29$	&	$-6.27$ $\pm$	 $0.05$\\

$071010A$	&	$0.980$	&	$A$	&	$514.15$		&	$-11.50$	$\pm$	$0.04$	&	$-7.53$	$\pm$	$0.03$	&	$0.87$	$\pm$	$0.09$	&	$-0.76$	$\pm$	$0.01$	&	$-7.29$	$\pm$	$0.42$	&	$-6.33$ $\pm$ 	$0.09$\\

$091024$	&	$1.092$	&	$A$	&	$442.00$		&	$-10.78$	$\pm$	$0.01$	&	$-6.98$	$\pm$	$0.04$	&	$1.83$	$\pm$	$0.05$	&	$-1.13$	$\pm$	$0.06$	&	$-6.46$	$\pm$	$0.06$	&	$-4.88$ $\pm$ 	$0.02$ \\

$061007$	&	$1.262$	&	$P$	&	$57.90$		&	$-8.92$	$\pm$	$0.01$	&	$-7.22$	$\pm$	$0.01$	&	$5.84$	$\pm$	$1.89$	&	$-1.52$	$\pm$	$0.08$	&	$-5.40$	$\pm$	$0.02$	&	$-3.94$ $\pm$ 	$0.01$ \\

 $130420A$	&	$1.297$	&	$A$	&	$1778.28$		& $-11.53$	$\pm$	$0.04$	&	&	&	&	$-6.69$	$\pm$	$0.04$	&	$-5.11$	\\	
									
 $090530$	&	$1.300$	&	$A$	&	$60.48$		&	$-11.44$	$\pm$	$0.04$	&	$-7.90$	$\pm$	$0.05$	&	$0.90$	&	$-0.68$	$\pm$	$0.05$	&	$-6.35$	$\pm$	$0.11$	&	$-5.73$	$\pm$	$0.04$	\\

  $061121$	&	$1.315$	&	$P$	&	$71.26$		&	$-10.61$	$\pm$	$0.04$	&	$-8.98$	$\pm$	$0.04$	&	$1.83$	$\pm$	$0.04$	&	$-3.24$	$\pm$	$0.15$	&	$-5.44$	$\pm$	$0.01$	&	$-4.55$ $\pm$ 	$0.01$ \\

 $061121$	&	$1.315$	&	$A$	&	$151.20$		&	$-10.84$	$\pm$	$0.04$	&	$-6.96$	$\pm$	$0.04$	&	$1.44$	&	$-1.05$	$\pm$	$0.01$	&	$-5.44$	$\pm$	$0.01$	&	$-4.55$	$\pm$	$0.01$	\\

 $100901A$	&	$1.410$	&	$A$	&	$4579.20$	&	$-11.50$	$\pm$	$0.04$	&	$-6.69$	$\pm$	$0.03$	&	&	&	$-7.26$	$\pm$	$0.27$	&	$-5.47$	$\pm$	$0.06$	\\
					
  $110213A$	&	$1.460$	&	$A$	&	$248.00$		&	$-10.43$	&	$-6.14$	$\pm$	$0.01$	&	$0.80$	&	$-0.59$	$\pm$	$0.05$	&	$-6.00$	$\pm$	$0.08$	&	$-5.06$	$\pm$	$0.03$ \\		
 $060418$	&	$1.489$	&	$A(U)$	&	$127.00$		&	$-10.33$	$\pm$	$0.06$	&	$-7.51$	$\pm$	$0.05$	&	$2.28$	$\pm$	$0.33$	&	$-1.13$	$\pm$	$0.12$	&	$-6.11$	$\pm$	$0.04$	&	$-4.88$ $\pm$ 	$0.01$ \\

  $080330$	&	$1.512$	&	$A$	&	$432.90$	&	$-12.29$	$\pm$	$0.04$	&	$-8.37$	$\pm$	$0.02$	&	&	&	$-7.36$	$\pm$	$0.22$	&	$-6.36$	$\pm$	$0.08$	\\	
				
  $990123$	&	$1.600$	&	$P$	&	$49.88$		&	$-8.43$	$\pm$	$0.07$	&	$-6.75$	$\pm$	$0.07$	&	$3.72$	&	$-2.37$	$\pm$	$0.05$	&	&	$-3.21$	\\	
				
  $080603A$	&	$1.687$	&	$P?$	&	$136.00$		&	$-13.07$	$\pm$	$0.10$	&	$-11.38$	$\pm$	$0.12$	&	$1.17$	&	$-4.26$	&	&	$-5.73$	\\			
				
  $080603A$	&	$1.687$	&	$A(U)$	&	$2112.00$	&	$-12.20$	$\pm$	$0.01$	&	$-8.05$	$\pm$	$0.03$	&	$0.04$	$\pm$	$0.01$	&	$-0.54$	$\pm$	$0.11$	&	&	$-5.73$	\\
				
  $080928$	&	$1.692$	&	$A$	&	$2117.84$	&	$-11.80$	$\pm$	$0.02$	&	$-7.88$	$\pm$	$0.03$	&	$2.20$	$\pm$	$0.68$	&	$-0.11$	$\pm$	$0.02$	&	$-6.66$	$\pm$	$0.07$	&	$-5.40$	$\pm$ $0.03$	\\

  $120119A$	&	$1.728$	&	$A$	&	$766.80$		&	$-11.02$	$\pm$	$0.08$	&	$-7.56$	$\pm$	$0.06$	&	$0.44$	$\pm$	$0.06$	&	$-0.65$	$\pm$	$0.06$	&	$-5.93$	$\pm$	$0.02$	&	$-4.63$ $\pm$ $0.02$	\\

  $100906A$	&	$1.730$	&	$A(U)$	&	$125.35$	&	$-9.95$	$\pm$	$0.01$	&	$-7.38$	$\pm$	$0.01$	&	$0.46$	&	$-0.47$	$\pm$	$0.06$	&	$-6.12$	$\pm$	$0.02$	&	$-4.86$	$\pm$	$0.01$	\\

  $081008$	&	$1.968$	&	$P?$	&	$107.63$		&	$-10.66$	$\pm$	$0.01$	&	$-7.80$	$\pm$	$0.03$	&	$1.26$	$\pm$	$0.13$	&	$-0.72$	$\pm$	$0.03$	&	$-6.94^{+0.08}_{-0.07}$	&	$-5.25^{+0.05}_{-0.04}$	\\		

   $081203$	&	$2.050$	&	$A$	&	$345.00$		&	$-9.99$	$\pm$	$0.04$	&	$-6.58$	$\pm$	$0.04$	&	$1.18$	&	$-1.62$	$\pm$	$0.05$	&	$-6.43$	$\pm$	$0.06$	&	$-4.87$	$\pm$	$0.02$	\\

 $110205A$	&	$2.220$	&	$P$	&	$226.00$		&	$-11.82$	$\pm$	$0.03$	&	$-9.73$	$\pm$	$0.07$	&	$3.43$	$\pm$	$0.20$	&	$-2.92$	$\pm$	$0.26$	&	$-6.60$	$\pm$	$0.03$	&	$-4.70$ $\pm$ $0.02$	\\

  $110205A$	&	$2.220$	&	$A$	&	$1040.00$		&	$-10.67$	$\pm$	$0.03$	&	$-7.30$	$\pm$	$0.06$	&	$1.36$	$\pm$	$0.26$	&	$-1.58$	$\pm$	$0.02$	&	$-6.60$	$\pm$	$0.03$	&	$-4.70$ $\pm$ $0.02$	\\

  $120815A$	&	$2.359$	&	$A$	&	$499.50$	&	$-12.27$	$\pm$	$0.01$	&	$-8.69$	$\pm$	$0.01$	&	$0.15$	$\pm$	$0.03$	&	$-0.50$	$\pm$	$0.02$	&	$-6.77$	$\pm$	$0.22$	&	$-6.27$	$\pm$ $0.05$	\\

  $080310$	&	$2.427$	&	$A(U)$	&	$228.50$		&	$-11.86$	$\pm$	$0.02$	&	$-8.35$	$\pm$	$0.02$	&	$1.53$	$\pm$	$0.13$	&	$-0.65$	$\pm$	$0.09$	&	$-7.17^{+0.13}_{-0.12}$	&	$-5.57$	$\pm$	$0.03$	\\

  $090812$	&	$2.452$	&	$P$	&	$73.55$		&	$-11.10$	$\pm$	$0.03$	&	$-9.37$	$\pm$	$0.04$	&	$1.03$	$\pm$	$0.03$	&	$-1.10$	$\pm$	$0.03$	&	$-6.20$	$\pm$	$0.04$	&	$-4.93$ $\pm$ $0.02$	\\

 $050820A$	&	$2.615$	&	$P$	&	$267.44$		&	$-11.47$	$\pm$	$0.01$	&	$-9.63$	$\pm$	$0.03$	&	$0.59$	$\pm$	$0.48$	&	$-3.59$	$\pm$	$0.68$	&	$-6.43$	$\pm$	$0.12$	&	$-5.06$ $\pm$ $0.03$	\\

  $050820A$	&	$2.615$	&	$A(U)$	&	$426.83$		&	$-11.20$	$\pm$	$0.01$	&	$-7.48$	$\pm$	$0.03$	&	$3.21$	$\pm$	$0.69$	&	$-0.89$	$\pm$	$0.02$	&	$-6.43$	$\pm$	$0.12$	&	$-5.06$ $\pm$ $0.03$	\\

  $080210$	&	$2.641$	&	$A$	&	$345.60$		&	$-10.90$	$\pm$	$0.04$	&	$-7.48$	$\pm$	$0.04$	&	$3.66$	&	$-1.36$	$\pm$	$0.02$	&	$-6.83$	$\pm$	$0.11$	&	$-5.56$	$\pm$	$0.03$	\\

  $071031$	&	$2.692$	&	$A$	&	$1023.80$	&	$-12.71$	&	$-8.85$	$\pm$	$0.01$	&	$1.27$	$\pm$	$0.07$	&	$-0.19$	$\pm$	$0.03$	&	$-7.64^{+0.13}_{-0.12}$	&	$-5.96$	$\pm$	$0.06$	\\			

  $090726$	&	$2.710$	&	$A$	&	$514.80$		&	$-12.13$	$\pm$	$0.03$	&	$-8.90$	$\pm$	$0.02$	&	$1.03$	$\pm$	$0.05$	&	$-1.48$	$\pm$	$0.16$	&	$-7.42^{+0.14}_{-0.13}$	&	$-6.19^{+0.05}_{-0.04}$	\\
		
  $091029$	&	$2.750$	&	$A$	&	$410.30$		&	$-12.13$	$\pm$	$0.01$	&	$-7.91$	$\pm$	$0.01$	&	$0.47$	&	$-0.55$	$\pm$	$0.01$	&	$-6.90$	$\pm$	$0.05$	&	$-5.55$	$\pm$	$0.03$	\\

  $070411$	&	$2.950$	&	$A$	&	$432.00$	&	$-12.04$	$\pm$	$0.02$	&	$-7.68$	$\pm$	$0.04$	&	$0.31$	$\pm$	$0.12$	&	$-0.42$	$\pm$	$0.30$	&	$-7.03$	$\pm$	$0.07$	&	$-5.37$	$\pm$ $0.02$	\\

 $060607A$	&	$3.075$	&	$P?$	&	$212.04$		&	$-11.09$	$\pm$	$0.01$	&	$-8.17$	$\pm$	$0.02$	&	$1.87$	$\pm$	$0.54$	&	$-1.31$	$\pm$	$0.08$	&	$-6.82^{+0.08}_{-0.07}$	&	$-5.43^{+0.06}_{-0.05}$	\\	
	
$060526$	&	$3.221$	&	$A(U)$	&	$192.79$		&	$-11.80$	$\pm$	$0.09$	&	$-8.13$	$\pm$	$0.08$	&	$0.43$	&	$-0.84$	$\pm$	$0.12$	&	$-6.87$	$\pm$	$0.10$	&	$-5.73$	$\pm$	$0.06$	\\

  $080810$	&	$3.355$	&	$P?$	&	$126.14$	&	$-9.98$	$\pm$	$0.05$	&	$-7.11$	$\pm$	$0.03$	&	&	$-1.94$	$\pm$	$0.12$	&	$-6.63$	$\pm$	$0.05$	&	$-5.01$	$\pm$	$0.03$	\\		

  $090313$	&	$3.380$	&	$A$	&	$1312.80$		&	$-11.11$	$\pm$	$0.04$	&	$-6.98$	$\pm$	$0.06$	&	$1.19$	$\pm$	$0.15$	&	$-1.31$	$\pm$	$0.13$	&	$-7.46$	$\pm$	$0.37$	&	$-5.67^{+0.07}_{-0.08}$	\\

  $060605$	&	$3.773$	&	$A$	&	$463.00$	&	$-10.38$	$\pm$	$0.09$	&	$-6.83$	$\pm$	$0.12$	&	$0.31$	$\pm$	$0.08$	&	$-0.84$	$\pm$	$0.04$	&	$-7.44^{+0.24}_{-0.23}$	&	$-6.25^{+0.07}_{-0.09}$	\\			

  $081029$	&	$3.848$	&	$A$	&	$5936.80$		&	$-12.29$	$\pm$	$0.01$	&	$-7.89$	$\pm$	$0.01$	&	$1.83$	$\pm$	$0.15$	&	$-0.67$	$\pm$	$0.03$	&	$-7.66$	$\pm$	$0.24$	&	$-5.48$ $\pm$ $0.05$	\\

  $060210$	&	$3.913$	&	$A$	&	$393.00$		&	$-11.55$	$\pm$	$0.06$	&	$-7.92$	$\pm$	$0.05$	&	$1.15$	$\pm$	$0.64$	&	$-1.11$	$\pm$	$0.04$	&	$-6.46$	$\pm$	$0.08$	&	$-4.85$ $\pm$ $0.02$	\\

  $050730$	&	$3.968$	&	$A$	&	$726.89$		&	$-11.63$	$\pm$	$0.09$	&	$-7.62$	$\pm$	$0.06$	&	$0.88$	$\pm$	$0.25$	&	$-0.86$	$\pm$	$0.06$	&	$-7.20$	$\pm$	$0.11$	&	$-5.59$ $\pm$ $0.04$	\\

  $060206$	&	$4.048$	&	$A$	&	$3534.00$		&	$-12.04$	$\pm$	$0.01$	&	$-7.75$	$\pm$	$0.01$	&	$4.65$	$\pm$	$0.30$	&	$-1.12$	$\pm$	$0.07$	&	$-6.57^{+0.07}_{-0.06}$	&	$-5.98^{+0.04}_{-0.03}$	\\		

  $080129$	&	$4.350$	&	$A$	&	$514.59$	&	$-9.05$	$\pm$	$0.04$	&	$-4.80$	$\pm$	$0.04$	&	&	&	$-7.26^{+0.32}_{-0.31}$	&	$-5.72^{+0.06}_{-0.07}$	\\		
							
  $071025$	&	$5.200$	&	$A$	&	$499.40$	&	$-11.54$	$\pm$	$0.03$	&	$-8.14$	$\pm$	$0.05$	&	$0.25$	$\pm$	$0.02$	&	$-0.81$	$\pm$	$0.13$	&	$-6.81$	$\pm$	$0.05$	&	$-4.96$	$\pm$ $0.01$	\\

  $050904$	&	$6.295$	&	$A(U)$	&	$518.40$		&	$-10.04$	$\pm$	$0.12$	&	$-6.91$	$\pm$	$0.02$	&	$3.46$	&	$-4.09$	&	$-7.06$	$\pm$	$0.08$	&	$-4.91$	$\pm$	$0.02$	\\		\hline

\end{longtable}
\end{landscape}

\begin{landscape}
\begin{longtable}{cccccccccc}

\caption{Parameters of gamma-ray bursts in the proper frame of reference} \\

&&&&&&&&& \\
	$GRB$ & $z$ & $Sample$ & $T_{peak}$  & $logL_{R} $ & $logE_{R}$  & $T_{90}$ & $logL_{iso}$ & $logE_{iso}$ & $\alpha$ \\
&&&&&&&&& \\
           &        &              &    $s$                    &       $erg/s$                 &   $erg$                    &         $s$              &          $erg/s$              &        $erg$                      \\ 
&&&&&&&&& \\
\hline
&&&&&&&&& \\

  $130427A$	&	$0.340$	&	$P$	&	$10.68$		&	$48.42$	$\pm$	$0.01$	&	$49.71$	$\pm$	$0.04$	&	$242.33$	$\pm$	$1.87$	&	$52.38$	$\pm$	$0.01$	&	$53.66^{+0.13}_{-0.09}$	&	$0.99$	$\pm$	$0.08$	\\

  $130831A$	&	$0.479$	&	$A$	&	$543.60$	&	$46.20$	$\pm$	$0.01$	&	$48.89$	$\pm$	$0.01$	&	$23.43$	$\pm$	$0.36$	&	$51.00$	$\pm$	$0.04$	&	&	$0.06$	$\pm$	$0.05$	\\	
	
  $081007$	&	$0.530$	&	$A$	&	$91.40$		&	$45.57$	$\pm$	$0.01$	&	$48.29$	$\pm$	$0.03$	&	$3.63$	$\pm$	$0.17$	&	$50.30^{+0.21}_{-0.21}$	&	$50.89^{+0.18}_{-0.20}$	&	$0.65^{+1.02}_{-0.64}$	\\

  $060729$	&	$0.543$	&	$P$	&	$61.97$		&	$45.55$	$\pm$	$0.06$	&	$47.05$	$\pm$	$0.10$	&	$77.22$	$\pm$	$0.91$	&	$50.11$	$\pm$	$0.13$	&	$52.00^{+0.08}_{-0.48}$	&	$0.23$	$\pm$	$0.14$	\\

  $060729$	&	$0.543$	&	$A$	&	$297.38$	&	$45.20$	$\pm$	$0.04$	&	$77.22$	$\pm$	$0.91$	&	$50.11$	$\pm$	$0.13$	&	$52.00^{+0.08}_{-0.48}$	&	$0.23$	$\pm$	$0.14$	\\

  $111209A$	&	$0.677$	&	$A$	&	$403.15$	&	$45.63$	$\pm$	$0.04$	&	&	$2.77$	$\pm$	$0.20$	&	$50.95$	$\pm$	$1.77$	&	$51.97^{+0.41}_{-1.14}$	&	$0.57^{+0.48}_{-0.49}$	\\						
  $060904B$	&	$0.703$	&	$A$	&	$308.88$		&	$45.59$	$\pm$	$0.01$	&	$49.41$	$\pm$	$0.02$	&	$100.44$	$\pm$	$1.34$	&	$50.59^{+0.13}_{-0.11}$	&	$51.45^{+0.34}_{-0.08}$	&	$1.00^{+0.69}_{-0.59}$	\\		
		
  $080710$	&	$0.845$	&	$A$	&	$1322.75$		&	$45.832$	&	$49.35$	&	$75.35$	$\pm$	$5.43$	&	$50.57$	$\pm$	$0.14$	&	$52.69^{+0.22}_{-0.85}$	&	$0.74$	$\pm$	$0.24$	\\		
		
  $080319B$	&	$0.937$	&	$P$	&	$14.27$		&	$50.28$	$\pm$	$0.07$	&	$51.47$	$\pm$	$0.04$	&	$76.06$	$\pm$	$1.29$	&	$52.70$	$\pm$	$0.01$	&	$55.05^{+0.02}_{-0.10}$	&	$0.91$	$\pm$	$0.02$	\\

  $071010B$	&	$0.950$	&	$A$	&	$70.25$	&	$45.96$	$\pm$	$0.04$	&	$47.80$	$\pm$	$0.05$	&	$17.78$	$\pm$	$0.52$	&	$51.40^{+0.04}_{-0.03}$	&	$52.25^{+0.16}_{-0.04}$	&	$0.53^{+0.23}_{-0.22}$	\\		
			
  $070419A$	&	$0.971$	&	$A$	&	$246.55$		&	$45.24$	$\pm$	$0.05$	&	$48.41$	$\pm$	$0.07$	&	$81.83$	$\pm$	$4.5$	&	$49.49$	$\pm$	$0.29$	&	$51.26^{+0.18}_{-0.10}$	&	\\		
	
  $071010A$	&	$0.980$	&	$A$	&	$259.67$		&	$46.35$	$\pm$	$0.04$	&	$50.03$	$\pm$	$0.03$	&	$11.31$	$\pm$	$0.86$	&	$50.40$	$\pm$	$0.42$	&	$51.56^{+0.12}_{-0.66}$	&	\\		
	
  $091024$	&	$1.092$	&	$A$	&	$211.24$		&	$47.24$	$\pm$	$0.01$	&	$50.73$	$\pm$	$0.04$	&	$54.83$	$\pm$	$2.37$	&	$51.36$	$\pm$	$0.06$	&	$53.46^{+0.06}_{-0.50}$	&	$0.72$	$\pm$	$0.07$	\\

  $061007$	&	$1.262$	&	$P$	&	$25.59$		&	$49.40$	$\pm$	$0.01$	&	$51.07$	$\pm$	$0.01$	&	$33.118$	$\pm$	$0.23$	&	$52.57$	$\pm$	$0.02$	&	$54.83^{+0.02}_{-0.28}$	&	$1.04$	$\pm$	$0.03$	\\

  $130420A$	&	$1.297$	&	$A$	&	$774.17$	&	$46.71$	$\pm$	$0.04$	&	&	$50.00$	$\pm$	$2.11$	&	$51.31$	$\pm$	$0.04$	&	$52.72$	&	$0.88^{+0.68}_{-0.41}$	\\						
  $090530$	&	$1.300$	&	$A$	&	$26.30$		&	$46.81$	$\pm$	$0.04$	&	$49.99$	$\pm$	$0.05$	&	$17.72$	$\pm$	$0.50$	&	$51.65$	$\pm$	$0.11$	&	&	$0.43$	$\pm$	$0.16$	\\
	
  $061121$	&	$1.315$	&	$P$	&	$30.79$		&	$47.64$	$\pm$	$0.04$	&	$48.90$	$\pm$	$0.04$	&	$35.86$	$\pm$	$5.40$	&	$52.57$	$\pm$	$0.01$	&	$53.83^{+0.02}_{-0.54}$	&	$0.63$	$\pm$	$0.03$	\\

  $061121$	&	$1.315$	&	$A$	&	$65.33$		&	$47.41$	$\pm$	$0.04$	&	$50.92$	$\pm$	$0.04$	&	$35.86$	$\pm$	$5.40$	&	$52.57$	$\pm$	$0.01$	&	$53.83^{+0.02}_{-0.54}$	&	$0.63$	$\pm$	$0.03$	\\

  $100901A$	&	$1.410$	&	$A$	&	$1900.08$	&	$46.85$	$\pm$	$0.04$	&	$51.27$	$\pm$	$0.03$	&	$190.54$	$\pm$	$4.42$	&	$50.83$	$\pm$	$0.27$	&	$52.82^{+0.16}_{-0.14}$	&	$0.45^{+0.22}_{-0.23}$	\\	
		
  $110213A$	&	$1.460$	&	$A$	&	$100.81$		&	$47.98$	&	$51.87$	$\pm$	$0.01$	&	$17.53$	$\pm$	$1.41$	&	$52.12$	$\pm$	$0.08$	&	$53.13^{+0.04}_{-0.22}$	&	$0.18$	$\pm$	$11$	\\	
	
  $060418$	&	$1.489$	&	$A(U)$	&	$51.02$		&	$48.07$	$\pm$	$0.06$	&	$50.49$	$\pm$	$0.05$	&	$41.48$	$\pm$	$4.15$	&	$52.03$	$\pm$	$0.04$	&	$53.41^{+0.03}_{-0.51}$	&	$0.39$	$\pm$	$0.05$	\\

  $080330$	&	$1.512$	&	$A$	&	$172.34$	&	$46.01$	$\pm$	$0.04$	&	$49.53$	$\pm$	$0.02$	&	$26.31$	$\pm$	$0.39$	&	$50.81$	$\pm$	$0.22$	&	$52.14^{+0.29}_{-0.34}$	\\					
  $990123$	&	$1.600$	&	$P$	&	$19.18$		&	$50.08$	$\pm$	$0.07$	&	$51.15$	$\pm$	$0.07$	&	&	&	$54.60$	\\				
						
  $080603A$	&	$1.687$	&	$P?$	&	$50.61$		&	$45.62$	$\pm$	$0.10$	&	$48.16$	$\pm$	$0.05$	&	&	&	$52.12$	&	\\	
								
  $080603A$	&	$1.687$	&	$A(U)$	&	$785.89$	&	$46.49$	$\pm$	$0.01$	&	$50.20$	$\pm$	$0.03$	&	&	&	$52.12$	&	\\		
								
 $080928$	&	$1.692$	&	$A$	&	$786.72$	&	$46.84$	$\pm$	$0.02$	&	$50.33$	$\pm$	$0.03$	&	$105.83$	$\pm$	$4.52$	&	$51.62$	$\pm$	$0.07$	&	$52.92^{+0.06}_{-0.37}$	&	$0.27^{+0.12}_{-0.13}$	\\		
	
  $120119A$	&	$1.728$	&	$A$	&	$281.09$		&	$47.59$	$\pm$	$0.08$	&	$50.61$	$\pm$	$0.06$	&	$25.81$	$\pm$	$1.59$	&	$52.38$	$\pm$	$0.02$	&	$53.32^{+0.04}_{-0.03}$	&	$0.95^{+0.14}_{-0.13}$	\\
		
  $100906A$	&	$1.730$	&	$A(U)$	&	$45.92$	&	$48.71$	$\pm$	$0.01$	&	$50.82$	$\pm$	$0.01$	&	$42.80$	$\pm$	$0.25$	&	$52.18$	$\pm$	$0.02$	&	$53.50^{+0.02}_{-0.40}$	&	$0.34$	$\pm$	$0.04$	\\	

  $081008$	&	$1.968$	&	$P?$	&	$36.26$		&	$48.18$	$\pm$	$0.01$	&	$49.70$	$\pm$	$0.01$	&	$67.15$	$\pm$	$3.88$	&	$51.50^{+0.08}_{-0.07}$	&	$52.83^{+0.11}_{-0.07}$	&	$0.70^{+0.34}_{-0.31}$	\\		
		
  $081203$	&	$2.050$	&	$A$	&	$113.11$		&	$48.76$	$\pm$	$0.04$	&	$51.69$	$\pm$	$0.04$	&	$83.37$	$\pm$	$8.83$	&	$52.06$	$\pm$	$0.06$	&	$53.76^{+0.04}_{-0.61}$	&	$0.56$	$\pm$	$0.06$	\\

  $110205A$	&	$2.220$	&	$P$	&	$70.19$		&	$47.01$	$\pm$	$0.03$	&	$48.59$	$\pm$	$0.07$	&	$86.03$	$\pm$	$1.45$	&	$51.97$	$\pm$	$0.03$	&	$53.49^{+0.22}_{-0.04}$	&	$0.61^{+0.18}_{-0.17}$	\\		

  $110205A$	&	$2.220$	&	$A$	&	$322.98$		&	$48.16$	$\pm$	$0.03$	&	$51.02$	$\pm$	$0.04$	&	$86.03$	$\pm$	$1.45$	&	$51.97$	$\pm$	$0.03$	&	$53.49^{+0.22}_{-0.04}$	&	$0.61^{+0.18}_{-0.17}$	\\	
	
  $120815A$	&	$2.359$	&	$A$	&	$148.72$	&	$46.77$	$\pm$	$0.01$	&	$49.83$	$\pm$	$0.01$	&	$2.88$	$\pm$	$0.36$	&	$51.88$	$\pm$	$0.22$	&	$52.74^{+0.21}_{-0.96}$	\\					
  $080310$	&	$2.427$	&	$A(U)$	&	$66.67$		&	$47.01$	$\pm$	$0.02$	&	$49.98$	$\pm$	$0.02$	&	$105.60$	$\pm$	$1.10$	&	$51.49^{+0.13}_{-0.12}$	&	$53.31^{+0.11}_{-0.68}$	\\	
					
  $090812$	&	$2.452$	&	$P$	&	$21.31$		&	$48.15$	$\pm$	$0.03$	&	$49.50$	$\pm$	$0.04$	&	$28.90$	$\pm$	$4.43$	&	$52.48$	$\pm$	$0.04$	&	$53.89^{+0.04}_{-0.61}$	&	$0.71$	$\pm$	$0.06$	\\

  $050820A$	&	$2.615$	&	$P$	&	$73.99$		&	$47.58$	$\pm$	$0.01$	&	$49.15$	$\pm$	$0.02$	&	$66.31$	$\pm$	$0.10$	&	$52.31$	$\pm$	$0.12$	&	$53.89^{+0.08}_{-0.11}$	&	$0.83$	$\pm$	$0.11$	\\

  $050820A$	&	$2.615$	&	$A(U)$	&	$118.08$		&	$47.85$	$\pm$	$0.02$	&	$51.06$	$\pm$	$0.02$	&	$66.31$	$\pm$	$0.10$	&	$52.31$	$\pm$	$0.12$	&	$53.89^{+0.08}_{-0.11}$	&	$0.83$	$\pm$	$0.11$	\\

 $080210$	&	$2.641$	&	$A$	&	$94.92$		&	$48.08$	$\pm$	$0.04$	&	$50.73$	$\pm$	$0.04$	&	$12.05$	$\pm$	$1.20$	&	$51.93$	$\pm$	$0.11$	&	$53.06^{+0.05}_{-0.24}$	&	$0.25$	$\pm$	$0.12$	\\

  $071031$	&	$2.692$	&	$A$	&	$277.30$	&	$46.57$	&	$49.87$	$\pm$	$0.01$	&	$50.70$	$\pm$	$1.93$	&	$51.13^{+0.13}_{-0.12}$	&	&	\\									
  $090726$	&	$2.710$	&	$A$	&	$138.76$		&	$47.06$	$\pm$	$0.03$	&	$49.72$	$\pm$	$0.02$	&	$13.75$	$\pm$	$0.26$	&	$51.36^{+0.14}_{-0.13}$	&	$52.26^{+0.24}_{-0.13}$	&	\\	
				
  $091029$	&	$2.750$	&	$A$	&	$109.41$		&	$46.98$	$\pm$	$0.01$	&	$50.63$	$\pm$	$0.01$	&	$10.66$	$\pm$	$0.34$	&	$51.89$	$\pm$	$0.05$	&	$52.84^{+0.19}_{-0.06}$	&	$0.60^{+0.32}_{-0.30}$	\\	
	
  $070411$	&	$2.950$	&	$A$	&	$109.37$	&	$47.26$	$\pm$	$0.02$	&	$51.02$	$\pm$	$0.02$	&	$27.48$	$\pm$	$0.92$	&	$51.84$	&	$53.33^{+0.04}_{-0.25}$	&	$0.30$	$\pm$	$0.10$	\\	
		
  $060607A$	&	$3.075$	&	$P?$	&	$52.04$		&	$48.21$	$\pm$	$0.01$	&	$50.52$	$\pm$	$0.02$	&	$25.17$	$\pm$	$0.82$	&	$52.10^{+0.08}_{-0.07}$	&	$52.95^{+0.12}_{-0.07}$	&	$0.99^{+0.32}_{-0.34}$	\\	
			
  $060526$	&	$3.221$	&	$A(U)$	&	$45.67$		&	$47.47$	$\pm$	$0.09$	&	$50.51$	$\pm$	$0.08$	&	$70.01$	$\pm$	$0.95$	&	$52.09$	$\pm$	$0.10$	&	$53.02^{+0.06}_{-0.18}$	&	\\
			
  $080810$	&	$3.355$	&	$P?$	&	$28.97$	&	$49.53$	$\pm$	$0.05$	&	$51.76$	$\pm$	$0.03$	&	$104.05$	$\pm$	$1.17$	&	$52.37$	$\pm$	$0.05$	&	$53.95^{+0.08}_{-0.07}$	&	$0.69$	$\pm$	$0.12$	\\	

  $090313$	&	$3.380$	&	$A$	&	$299.73$		&	$48.36$	$\pm$	$0.04$	&	$51.85$	$\pm$	$0.06$	&	$20.60$	$\pm$	$1.54$	&	$51.56$	$\pm$	$0.37$	&	$53.12^{+0.07}_{-0.39}$	&	\\	
		
  $060605$	&	$3.773$	&	$A$	&	$97.00$	&	$49.46$	$\pm$	$0.09$	&	$52.33$	$\pm$	$0.12$	&	$3.88$	$\pm$	$0.24$	&	$51.69^{+0.24}_{-0.23}$	&	$52.22^{+0.36}_{-0.07}$	&	\\		
				
  $081029$	&	$3.848$	&	$A$	&	$1224.61$		&	$47.55$	$\pm$	$0.01$	&	$51.26$	$\pm$	$0.01$	&	$34.88$	$\pm$	$1.76$	&	$51.51$	$\pm$	$0.24$	&	$53.49^{+0.12}_{-0.10}$	&	$0.57$	$\pm$	$0.20$	\\

  $060210$	&	$3.913$	&	$A$	&	$79.99$		&	$48.11$	$\pm$	$0.06$	&	$51.05$	$\pm$	$0.05$	&	$75.30$	$\pm$	$4.20$	&	$52.72$	$\pm$	$0.08$	&	$54.10^{+0.05}_{-0.56}$	&	$0.53$	$\pm$	$0.10$	\\

  $050730$	&	$3.968$	&	$A$	&	$146.31$		&	$47.94$	$\pm$	$0.09$	&	$50.85$	$\pm$	$0.06$	&	$12.17$	$\pm$	$0.45$	&	$51.98$	$\pm$	$0.11$	&	$53.41^{+0.08}_{-0.60}$	&	$0.61$	$\pm$	$0.13$	\\

  $060206$	&	$4.048$	&	$A$	&	$700.08$		&	$47.68$	$\pm$	$0.01$	&	$51.27$	$\pm$	$0.01$	&	$1.20$	$\pm$	$0.03$	&	$52.64^{+0.07}_{-0.06}$	&	$52.61^{+0.07}_{-0.05}$	&	$0.82^{+0.33}_{-0.31}$	\\	
			
  $080129$	&	$4.350$	&	$A$	&	$96.18$	&	$50.62$	$\pm$	$0.04$	&	$54.14$	$\pm$	$0.04$	&	$8.52$	$\pm$	$0.56$	&	$52.01^{+0.32}_{-0.31}$	&	$53.42^{+0.15}_{-0.57}$	&	$0.78^{+0.253}_{-0.24}$	\\	
				
  $071025$	&	$5.200$	&	$A$	&	$80.55$	&	$48.76$	$\pm$	$0.03$	&	$51.36$	$\pm$	$0.05$	&	$26.00$	$\pm$	$0.87$	&	$52.64$	$\pm$	$0.05$	&	&	$0.33$	$\pm$	$0.06$	\\		
  $050904$	&	$6.295$	&	$A(U)$	&	$71.06$		&	$50.46$	$\pm$	$0.12$	&$52.73$	$\pm$	$0.12$	&	$27.03$	$\pm$	$0.31$	&	$52.59$	$\pm$	$0.08$	&	$54.38^{+0.04}_{-0.24}$	&	$0.83$	$\pm$	$0.07$	\\

\hline

\end{longtable}
\end{landscape}

\begin{longtable}{ccccccccc}

\caption{Pair correlations between  gamma-ray burst parameters with correlation coefficients $R>0.5$ and significance levels
${\rm SL}<1\%$.  The table lists the linear regression parameters ($a + bx$) (the variables are given in the logarithmic scale)}\\

\hline

& Correlation & Sample & $N$ & $R$ & $SL$ &  $a$  &  $b$  \\

\hline
    &&P+P? &11&0.86& $0.72 \cdot 10^{-3}$ &$-19.44 \pm 16.25$ &$1.40 \pm 0.31$ 	\\
         &&A&30&0.80& $0.91 \cdot 10^{-7}$ & $11.00 \pm 5.85$ & $0.81 \pm 0.11$     \\
          &$E_{iso}-L_{iso}$&A+A(U)+P+P?&47&0.84& $0.11 \cdot 10^{-12}$ &$4.37 \pm 4.64$ &$0.94 \pm 0.09$ \\
&&P&8&0.89&$ 0.32 \cdot 10^{-2}$ &$3.29 \pm 10.67$ &$0.97 \pm 0.20$ \\
&&A+A(U)&36&0.82&$ 0.86 \cdot 10^{-9}$ &$8.32 \pm 5.31$ &$0.86 \pm 0.10$ \\
&&A+A(U)+P?&39&0.82&$ 0.13 \cdot 10^{-10}$ &$8.10 \pm 5.09$ &$0.87 \pm 0.10$  \\

\hline
&&P+P? & 13 & 0.95 & $0.51 \cdot 10^{-6} $& $6.81 \pm 4.14$ &$ 0.89 \pm 0.09$ \\
&&A & 32 & 0.92 & $0.42 \cdot 10^{-13}$ &$ 3.82 \pm 3.51$ & $0.99 \pm 0.07$  \\
&&A(U)&7&0.92& $0.29 \cdot 10^{-2}$ & $19.67 \pm 5.76$ & $0.65 \pm 0.12$\\
&$E_{R}-L_{R}$&A+A(U)+P+P?&52&0.80&$0.92 \cdot 10^{-12}$ & $13.13 \pm 3.92$ & $0.78 \pm 0.08$ \\
&&A+A(U)&39&0.91&$0.89 \cdot 10^{-15}$ & $7.80 \pm 3.10$ & $0.90 \pm 0.07$ \\
&&P&9&0.99& $0.11 \cdot 10^{-6}$ & $4.74 \pm 2.06$ & $0.93 \pm 0.04$\\
&&A+A(U)+P?&43&0.89& $0.13 \cdot 10^{-14}$ & $9.02 \pm 3.31$ & $0.87 \pm 0.04$  \\
\hline

&&P+P?&13&0.76&$ 0.25 \cdot 10^{-2}$ & $-10.75 \pm 15.56$ & $1.13 \pm 0.29$ \\
&&A&32&0.65& $0.16 \cdot 10^{-3}$ & $-7.39 \pm 13.13$ & $1.10 \pm 0.50$ \\
&$E_{R}-E_{iso}$&A+A(U)+P+P?&48&0.52& $0.14 \cdot 10^{-3}$ & $-9.10 \pm 9.95$ & $0.78 \pm 0.19$ \\
&&A+A(U)&35&0.66& $0.15 \cdot 10^{-4}$ & $-4.02 \pm 10.79$ & $1.03 \pm 0.20$ \\
&&P&9&0.96& $0.42 \cdot 10^{-4}$ & $-32.27 \pm 9.07$ & $1.52 \pm 0.17$ \\
&&A+A(U)+P?&39&0.68& $0.22 \cdot 10^{-5}$ & $-7.07 \pm 10.29$ & $1.09 \pm 0.19$ \\
\hline

&&A&34&0.79& $0.30 \cdot 10^{-7}$ & $44.68 \pm 0.37$ & $5.39 \pm 0.74$ \\
&$L_{R}-z+1$&A+A(U)+P+P?&54&0.56& $0.13 \cdot 10^{-4}$ & $45.48 \pm 0.45$ & $4.36 \pm 0.90$ \\
&&A+A(U)&41&0.78& $0.19 \cdot 10^{-8}$ & $44.67 \pm 0.36$ & $5.54 \pm 0.71$ \\
&&A+A(U)+P?&45&0.77& $0.54 \cdot 10^{-9}$ & $44.58 \pm 0.37$ & $5.57 \pm 0.72$ \\

\hline

&&A&32&0.65& $0.65 \cdot 10^{-4}$ & $-4.92 \pm 11.95$ & $1.07 \pm 0.23$ \\
&$E_{R}-L_{iso}$&A+A(U)+P+P?&49&0.56& $0.30 \cdot 10^{-4}$ & $1.76 \pm 10.51$ & $0.94 \pm 0.20$ \\
&&A+A(U)&38&0.66& $0.66 \cdot 10^{-5}$ & $-5.49 \pm 10.63$ & $1.09 \pm 0.21$ \\
&&A+A(U)+P?&41&0.66& $0.23 \cdot 10^{-5}$ & $-5.79 \pm 10.17$ & $1.09 \pm 0.20 $ \\

\hline

&&P+P?&13&0.87&$ 0.12 \cdot 10^{-3}$ & $-25.60 \pm 12.69$ & $1.37 \pm 0.24$ \\
&&A&30&0.68& $0.40 \cdot 10^{-4}$ & $-9.22 \pm 11.60$ & $1.06 \pm 0.22$\\
&$L_{R}-E_{iso}$&A+A(U)+P+P?&50&0.79& $0.15 \cdot 10^{-10}$ & $-16.12 \pm 7.24$ & $1.20 \pm 0.14$ \\
&&A+A(U)&37&0.72& $0.40 \cdot 10^{-6}$ & $-13.66 \pm 9.81$ & $1.15 \pm 0.19$\\
&&P&9&0.94& $0.15 \cdot 10^{-3}$ & $-37.50 \pm 11.53$ & $1.59 \pm 0.21$ \\
&&A+A(U)+P?&41&0.74& $0.34 \cdot 10^{-7}$ & $-17.09 \pm 9.41$ & $1.22 \pm 0.18$ \\

\hline

&&A&34&0.74& $0.51 \cdot 10^{-6}$ & $-13.49 \pm 9.60$ & $1.18 \pm 0.19$ \\
&$L_{R}-L_{iso}$&A+A(U)+P+P?&51&0.78& $0.21 \cdot 10^{-10}$ & $-20.33 \pm 7.80$ & $1.31 \pm 0.15$ \\
&&A+A(U)&40&0.76& $0.14 \cdot 10^{-7}$ & $-18.38 \pm 9.10$ & $1.27 \pm 0.18$\\
&&A+A(U)+P?&43&0.76& $ 0.30 \cdot 10^{-8}$ & $-20.36 \pm 8.90$ & $1.31 \pm 0.18$ \\

\hline

&&A&30&0.59& $0.59 \cdot 10^{-3}$ & $51.49 \pm 0.35$ & $2.75 \pm 0.71$ \\
&$E_{iso}-z+1$&A+A(U)&37&0.62& $0.79 \cdot 10^{-2}$ & $51.47 \pm 0.32$ & $2.93 \pm 0.63$\\
&&A+A(U)+P?&41&0.63& $0.12 \cdot 10^{-4}$ & $51.43 \pm 0.30$ & $3.01 \pm 0.60$ \\

\hline

&$L_{R}-T{peak}$&P+P?&13&-0.69&$0.01$&$53.94 \pm 1.87$ & $-3.85 \pm 1.22$\\

\hline

&&A&32&0.72&$0.34 \cdot 10^{-5}$ & $47.97 \pm 0.47$ & $5.29 \pm 0.93$\\
&$E_{opt}-z+1$&A+A(U)+P+P?&52&0.63& $0.50 \cdot 10^{-6}$ & $48.06 \pm 0.42$ & $4.84 \pm 0.84$ \\
&&A+A(U)&39&0.73& $0.12 \cdot 10^{-6}$ & $48.01 \pm 0.41$ & $5.18 \pm 0.79$ \\
&&A+A(U)+P?&43&0.71& $0.76 \cdot 10^{-7}$ & $47.89 \pm 0.42$ & $5.28 \pm 0.84$ \\
\hline

&&A&34&0.69& $0.56 \cdot 10^{-5}$ & $50.11 \pm 0.27 $ & $2.98 \pm 0.55$ \\
&$L_{iso}-z+1$&A+A(U)+P+P?&51&0.55& $0.29 \cdot 10^{-4}$ & $50.57 \pm 0.26$ & $2.41 \pm 0.52$ \\
&&A+A(U)&40&0.69& $0.70 \cdot 10^{-6}$ & $50.18 \pm 0.25$ & $2.92 \pm 0.49$ \\
&&A+A(U)+P?&43&0.70& $0.14 \cdot 10^{-6}$ & $50.17 \pm 0.24$ & $2.98 \pm 0.47$ \\

\hline

\end{longtable}

\end{document}